\documentclass[10pt,fleqn]{article}
\usepackage{psfrag}
\usepackage{tikz}
\usepackage{pgfplots}
\usepackage{subcaption}
\usetikzlibrary{positioning}
\pgfplotsset{compat=newest} 
\pgfplotsset{plot coordinates/math parser=false} 
\usepackage{amsmath}
\usepackage[dvips]{epsfig}
\usepackage{epsfig}
\usepackage{epstopdf}
\usepackage{setspace}
\usepackage{amsmath,amssymb,bm}
\usepackage{graphics}
\usepackage{booktabs}
\usepackage[pdftex,colorlinks,bookmarks,pdfpagemode=UseOutlines,linkcolor=black,
pagecolor=black,urlcolor=black,citecolor=black,letterpaper,pageanchor=false]{hyperref}
\usepackage[margin=1in]{geometry}
\usepackage[T1]{fontenc}
\usepackage{titlesec}
\usepackage[pagewise,mathlines]{lineno}
\usepackage{amsthm}
\usepackage{longtable}
\usepackage{xfrac}
\usepackage{subcaption}
\usepackage{multirow}
\usepackage{hhline}

\newtheorem{theo}{Theorem}[section]
\newtheorem{lemma}{Lemma}[section]
\newtheorem{df}{Definition}[section]
\newtheorem{cor}{Corollary}[section]
\newtheorem{assump}{Assumption}[section]
\newtheorem{assert}{Assertion}[section]
\newtheorem{remark}{Remark}[section]
\newcommand{\bl}{\begin{lemma}}
\newcommand{\el}{\end{lemma}}
\newcommand{\be}{\begin{equation}}
\newcommand{\ee}{\end{equation}}
\newcommand{\beqn}{\begin{eqnarray}}
\newcommand{\eeqn}{\end{eqnarray}}
\newcommand{\bt}{\begin{theo}}
\newcommand{\et}{\end{theo}}
\newcommand{\bd}{\begin{df}}
\newcommand{\ed}{\end{df}}
\newcommand{\ba}{\begin{assump}}
\newcommand{\ea}{\end{assump}}
\newcommand{\bass}{\begin{assert}}
\newcommand{\eass}{\end{assert}}
\newcommand{\brem}{\begin{remark}}
\newcommand{\erem}{\end{remark}}
\newcommand{\bc}{\begin{cor}}
\newcommand{\ec}{\end{cor}}

\numberwithin{equation}{section}

\long\def\comment#1{}

\title{Modeling and Measuring Impact of Adaptive and Cooperative Adaptive Cruise Control
on Throughput of Signalized Intersections}
\author{Armin Askari, Daniel Albarnaz Farias, Alex A. Kurzhanskiy, Pravin Varaiya}

\begin{document}
\maketitle

\begin{abstract}
To properly assess the impact of (cooperative) adaptive cruise control
ACC (CACC), one has to model vehicle dynamics.
First of all, one has to choose the car following model, as it determines
the vehicle flow as vehicles accelerate from standstill or decelerate
because of the obstacle ahead.
The other factor significantly affecting the intersection throughput
is the maximal vehicle acceleration rate.

In this paper, we analyze three car following behaviors: Gipps model,
Improved Intelligent Driver Model (IIDM) and Helly model.
Gipps model exhibits rather aggressive acceleration behavior.
If used for the intersection throughput estimation, this model
would lead to overly optimistic results.
Helly model is convenient to analyze due to its linear nature,
but its deceleration behavior in the presence of obstacles ahead
is unrealistically abrupt.
Showing the most realistic acceleration and deceleration behavior of
the three models, IIDM is suited for ACC/CACC impact evaluation better
than the other two.

We discuss the influence of the maximal vehicle acceleration rate and
presence of different portions of ACC/CACC vehicles on intersection 
throughput in the context of the three car following models.
The analysis is done for two cases:
(1) free road downstream of the intersection; and
(2) red light at some distance downstream of the intersection.

Finally, we introduce the platoon model
and evaluate ACC and CACC with platooning in terms of travel time
using SUMO simulation of a traffic network with 7 intersections
in North Bethesda, MD.
\end{abstract}

{\bf Keywords}: signalized intersections, adaptive cruise control (ACC), 
cooperative adaptive cruise control (CACC), car following models,
Helly Model, Gipps Model, Improved Intelligent Driver Model

\section{Introduction and Background}\label{sec_intro}
Urban transportation is heading towards a crisis.
According to the 2015 Urban Mobility Scorecard~\cite{umsc15},
travel delays due to traffic congestion caused drivers to waste more than
3 billion gallons of fuel and kept travelers stuck in their cars for nearly
7 billion extra hours --- 42 hours per rush-hour commuter.
The total nationwide price tag: \$160 billion, or \$960 per commuter.
Almost two thirds of congestion in large cities (and more than 80\%
in smaller urban areas) occur on city streets.

Intersections are the major bottlenecks of the city road networks because 
an intersection's capacity is only a fraction of the vehicle flows that the 
streets connecting to the intersection can carry.
Traditional way of controlling traffic on city streets
is through adjusting signal timing.
There is a substantial body of work addressing signal optimization.
Tuning of cycle length is discussed in~\cite{kesur12,baietal12}.
Route prioritization through split adjustment is described 
in~\cite{smithetal15}.
The max-pressure local feedback control policy that gives priority to
movements with larger queues is also working with splits~\cite{maxpressure}.
Bandwidth maximization through offsets problem was solved in different
variations in~\cite{morganlittle64, little66, gartneretal91, papolafusco98, pillaietal98, gartneretal02}.
This work was analyzed and a new formulation 
relying on the concepts of relative offset and vehicle
arrival functions together with an efficient computation
technique was proposed in~\cite{gomes15}.
These results were extended in~\cite{nunzio_BOVSLCDC2015}
to optimize energy consumption via signal offsets control
and variable speed limits.
A review of adaptive signal systems is given in~\cite{abbasetal01} and
Traffic-responsive Urban Control (TUC) strategy is described 
in~\cite{diakaki2003}.
There is also an active research on improving intersection performance by
fully shifting the traffic control to vehicles~\cite{azimietal13, azimietal15}.
These approaches, however, are based on the assumption that all vehicles
can communicate with each other, which at present is unrealistic.

It is argued in~\cite{kurvar15} that as a combination of
efficient signal control, proper utilization of new vehicle technologies,
multimodal transportation; and parking management.
This paper presents an empirical study of how the
existing technologies, such as \emph{adaptive cruise control} (ACC),
and the emerging technologies,
such as \emph{cooperative adaptive cruise control}
(CACC)~\cite{milanes14}, can mitigate the congestion problem on city streets
by increasing the throughput of intersections, as vehicles equipped with 
ACC/CACC can travel closer to each other and even form \emph{platoons}.
Platoons were shown to improve freeway throughput~\cite{varaiya93},
and recently in~\cite{fernandes-nunes15}.
In~\cite{liorisetal15} the authors suggest that the saturation flow
rates, and hence intersection capacity, can be [at least] doubled by platooning.
However, that argument rests on the results of Point Queue modeling
of signalized arterials, which do not take into
account vehicle dynamics and car following behavior.

To better assess the impact of ACC and CACC vehicles on intersection
throughput, one has to model car following behavior.
The choice of the car following models largely determines the findings of
of the ACC/CACC impact study.
There is a variety of car following models.
The list starts with the Reuschel~\cite{reuschel50} and Pipes~\cite{pipes53}
models, in which the speed changes instantaneously as a function of
the distance to the leading vehicle.
Another class of models, generally referred to as Gazis-Herman-Rothery
(GHR)~\cite{gm58, ghr61},\footnote{Also known as
General Motors car following model.}
is where the acceleration depends on speed difference and the distance
gap according to the power law and not influenced by the driver's own
speed.
These models are incomplete in the following sense:
they can describe either free traffic or approach to standing obstacle,
but not both.
On the contrary, \emph{complete} models describe all situations, including
acceleration and cruising in free traffic, following other vehicles in
stationary and non-stationary situations, and approaching slow or standing
vehicles, and red lights~\cite{treiberkesting_book} (Chapter 10).

The class of models, where the acceleration depends on speed difference
with car in front and on the difference between the actual and the desired
gap linearly, is attributed to Helly~\cite{helly59}.
Helly model is complete, and
its linear nature makes it easy to understand and analyze.
It was extensively studied, built upon and used for ACC/CACC
modeling~\cite{swaroop94, godboleetal99, vanderwerfetal02, hoogendoornetal06, vanaremetal06}.

Another example of a complete model is the
Optimal Velocity Model (OVM)~\cite{bandoetal95}.
In OVM, acceleration
depends only on the distance (but not on the speed difference)
to the car in front: this distance determines
the optimal speed, which the vehicle tries to hold.
OVM is not always collision-free.
Full Velocity Difference Model (FVDM)~\cite{jiangetal01} extends OVM
by adding the linear dependence on the speed difference with the car
in front to the acceleration equation.
Generally considered more realistic than OVM, in terms of acceleration
values and the shock waves that it produces, FVDM suffers from the defect
that it is not complete in the sense defined above.
The reason is that the speed difference term does not depend on
the gap between the vehicle and the car in front.
Consequently, a slow vehicle triggers a significant deceleration of its
follower even if it is miles away.
Newell car following model~\cite{newell61, newell02},
describes car-following behaviour based on the analysis of time-space
trajectory, assuming that the time-space trajectory for 
two adjacent vehicles is essentially the same, except for
the shift in time and space.
For gaps smaller than desired and triangular fundamental diagram, the
Newell model behaves the same OVM.

All the above mentioned models are heuristic --- they attempt to describe
vehicle flow based on observations and common sense.
Gipps car following model~\cite{gipps81} and the Intelligent Driver Model
(IDM)~\cite{treiberetal2000} are similar to complete heuristic models
in that they too are defined by their acceleration equations.
In addition to that, they adhere the following principles:
\begin{enumerate}
\item the model is complete in the sense of the definition above;
\item the equilibrium gap to the car in front is no less than
the \emph{safe distance} computed as a sum of the minimal gap and the distance
the car can travel during the period called reaction time;
\item deceleration increases and decreases smoothly under normal driving
conditions, but can exceed ``comfortable'' level when the car in front is too
close and too slow --- to avoid collision;
\item transitions between different driving modes are smooth;
\item each model parameter describes only one aspect of driving behavior;
\item acceleration is strictly decreasing function of the speed;
\item acceleration is an increasing function of the gap between the vehicle
and the car in front;
\item acceleration is an increasing function of the speed of the car in front;
\item minimal gap to the car in front is maintained even during standstill,
but there is no backward movement if the gap is smaller than the minimal
(e.g. due to initial conditions).
\end{enumerate}
Gipps model is widely used and is implemented in the Aimsun microsimulator.
Krauss car following model~\cite{krauss98} is a stochastic variation
of Gipps model, where an auto-correlated noise is added to the vehicle
speed.
Krauss-Gipps model is implemented in the SUMO microsimulator~\cite{sumo}.
IDM is considered to have more realistic acceleration profile than that
of Gipps model.
It is widely investigated for the purpose of ACC/CACC
implementation~\cite{kestingetal10, schakeletal10, wangetal13, milanes14}.
Due to the continuous transition between free flow and congested traffic,
the gap between vehicles grows infinitely large as the speed approaches the
equilibrium value.
The other effect of this continuous transition is that the gap size between
the vehicle and the car in front smaller than desired leads to unrealistically
high deceleration values.
This produces unacceptable vehicle behavior in platoons.
Therefore, the acceleration function was modified to retain the spirit of IDM
but to eliminate its shortcomings.
The resulting model is called Improved IDM (IIDM)~\cite{treiberkesting_book}
(Chapter 11).
Gipps model and IIDM are sometimes called first \emph{principle models},
referring to the four principles stated above.

Finally, there is a category of car following models referred to as
\emph{behavioral}.
These are Wiedemann~\cite{wiedemann74, wiedemannreiter92}
and Fritzsche~\cite{fritzsche94} models, implemented in PTV Vissim and
Paramics simulators respectively.
These two models are essentially hybrid systems, where guards between different
modes of vehicle dynamics are thresholds for the speed difference and
the distance to the car in front.
Adjusting those guards changes driver behavior in the range from overly
cautious to aggressive and from slow reactive to fast.
In the Wiedemann and Fritzsche models, transitions between different driving
modes are not necessarily smooth, as acceleration changes in a series of
discrete transitions, which violates rule 4 of the first principles listed
above.
Designed to model human behavior and requiring complex tuning of the multitude
of parameters, these models are generally not used for ACC/CACC.

In this paper we analyze three models --- Gipps, IIDM and Helly --- and
assess the impact of ACC and CACC vehicles on the intersection throughput
in the context of these models.
The rest of the paper is organized as follows.
In Section~\ref{sec_cfm}, we review Gipps model, IIDM and Helly model and
analyze their properties in the context of signalized arterials.
In Section~\ref{sec_micro_macro}, we derive the \emph{macroscopic} models
corresponding the three car following models in question.
In Section~\ref{sec_acc_cacc}, we investigate the impact of ACC- and
CACC-enabled traffic on intersection throughput.
In Section~\ref{sec_platoon_model} we introduce the platoon model,
and evaluate it in the simulation of North Bethesda road network,
a case study presented in Section~\ref{sec_simulation}.
Section~\ref{sec_conclusion} concludes the paper.

\section{Analysis of Car Following Models}\label{sec_cfm}
We start by introducing the notation that will be used in the car following
model discussion.
It is summarized in Table~\ref{tab-micro-notation}
together with the default
parameter values that we will use in our experiments, unless stated otherwise.

\begin{table}[h]
\begin{center}
\begin{tabular}{|l|l|l|}
\hline
Symbol & Description & Default value\\
\hline
$l$ & Vehicle length. & $l=5$ m.\\
$t$, $\Delta t$ & Time and the model time step. & $\Delta t=0.05$ s.\\
$x(t)$ & Vehicle position. &\\
$x_l(t)$ & Position  of the vehicle in front, the \emph{leader}.&\\
$v_{\max}$ & Maximal admissible speed for the vehicle. & $v_{\max}=20$ m/s.\\
$v(t)$ & Vehicle speed. &\\
$v_l(t)$ & Speed of the leader. &\\
$a(t)$ & Vehicle acceleration. &\\
$a_{\max}$ & Maximal vehicle acceleration. & $a_{\max}=1.5 \mbox{ m/s}^2$. \\
$b$ & Desired vehicle deceleration. & $b=2 \mbox{ m/s}^2$.\\
$g(t)$ & \begin{tabular}{@{}l@{}}Gap: distance from the front of the vehicle
to the tail of the leader,\\
$g(t)=x_l(t)-x(t)-l$.\end{tabular} &\\
$g_{\min}$ & Minimal gap that is allowed between the
vehicle and the leader. & $g_{\min}=4$ m.\\
$g_d(t)$ & Desired gap between the vehicle and the
leader. & \\
$\tau$ & Reaction time to decelerate for the vehicle
to avoid collision with the leader. & $\tau=2.05$ s.\\
$\theta(t)$ & Headway: $\theta(t) = \frac{x_l(t)-x(t)}{v(t)}$. &\\
$f(t)$ & Vehicle flow: $f(t) = \frac{1}{\theta(t)}$. &\\
\hline
\end{tabular}
\end{center}
\caption{Notation for car following models.}
\label{tab-micro-notation}
\end{table}

The state update equations for a car following model are:
\begin{eqnarray}
v(t+\Delta t) & = & v(t) + a(t)\Delta t ; \label{eq_vehicle_speed}\\
x(t+\Delta t) & = & x(t) + \frac{v(t)+v(t+\Delta t)}{2}\Delta t = 
x(t) + v(t)\Delta t + \frac{a(t)\Delta t^2}{2}, \label{eq_vehicle_position}
\end{eqnarray}
where acceleration $a(t)$ depends on the car following model --- Gipps,
IIDM and Helly --- whose descriptions follow.

\begin{itemize}
\item Gipps model:
\be
a(t) = \min\left\{a_{\max}, \frac{v_{\max} - v(t)}{\Delta t},
\frac{1}{\Delta t}\left(-v(t) - b\tau +
\sqrt{(b\tau)^2 + (v_l(t))^2 + 2b\left(g(t) - g_{\min}\right)} \right)\right\}.
\label{eq_gipps}
\ee
\item IIDM:
\be
a(t) = \left\{\begin{array}{ll}
a_{\max}\left(1-\left(\frac{g_d(t)}{g(t)}\right)^{\delta_1}\right), &
\mbox{ if } \frac{g_d(t)}{g(t)} > 1;\\
a^\ast(t)\left(1-
\left(\frac{g_d(t)}{g(t)}\right)^{\delta_1 a_{\max}/a^\ast(t)}\right), &
\mbox{ otherwise,}
\end{array}\right.
\label{eq_iidm}
\ee
where
\begin{eqnarray}
a^\ast(t) & = & a_{\max}\left(1-
\left(\frac{v(t)}{v_{\max}}\right)^{\delta_2}\right),
\label{eq_iidm_afree}\\
g_d(t) & = & g_{\min} + \max\left\{0, v(t)\tau +
\frac{v(t)(v(t)-v_l(t))}{2\sqrt{a_{\max}b}}\right\}, \label{eq_iidm_desired_gap}
\end{eqnarray}
and $\delta_1, \delta_2$ are some fixed positive parameters.\footnote{In
the original descriptions of IDM and IIDM, parameter $\delta_1$ at
$\delta_1=2$,
but we believe, it does not have to be restricted to a given value.}

\item Helly model:
\be
a(t) = \min\left\{a_{\max}, \frac{v_{\max} - v(t)}{\Delta t},
\alpha_1\left(v_l(t) - v(t)\right) + 
\alpha_2\left(g(t) - g_{\min} - v(t)\tau\right)
\right\}
\label{eq_helly},
\ee
where $\alpha_1$ and $\alpha_2$ are
some positive fixed parameters.\footnote{Helly determined that $\alpha_1$
should be in the range $[0.17, 1.3]$, and $\alpha_2$ could be selected
in the range
$\left[\frac{1}{4}\alpha_1, \frac{1}{2}\alpha_1\right]$ \cite{brackstone99}.}
\end{itemize}

In all three models, the equilibrium speed ($a(t)=0$)
is achieved with $v(t) = v_l(t) = v_{\max}$
and $g(t) = g_{\min} + v(t)\tau$.
In this case, equilibrium headway is:
\be
\theta_e = \tau + \frac{g_{\min} + l}{v_{\max}}.
\label{eq_min_headway}
\ee
Assuming default values from Table~\ref{tab-micro-notation},
the length of the car $l=5$ m, the speed limit $v_{\max}=20$ m/s,
the minimal gap $g_{\min}=4$ m and the reaction time $\tau=2.05$ s, 
from~\eqref{eq_min_headway} we get $\theta_e=2.5$ s,\footnote{Empirical
evidence suggests that 2.5 seconds is a typical headway observed in dense
traffic on urban streets.}
which translates
to $f_e=1/\theta_e=0.4$ vehicles per second, 24 vehicles per minute,
or 1440 vehicles per hour.

Now let us compare the behavior of these car following models in three
experiments with intersections.
In these experiments we will be using the parameter values from
Table~\ref{tab-micro-notation}, and specific IIDM and the
Helly model parameters given in Table~\ref{tab-model-defaults}.

\begin{table}[h]
\begin{center}
\begin{tabular}{|l|}
\hline
$\delta_1 = 8$ \\
$\delta_2 = 4$ \\
$\alpha_1 = 0.5$ \\
$\alpha_2 = 0.25$ \\
\hline
\end{tabular}
\end{center}
\caption{Parameter values used for IIDM and the Helly model.}
\label{tab-model-defaults}
\end{table}

\begin{figure}
\centering
\includegraphics[width=\textwidth]{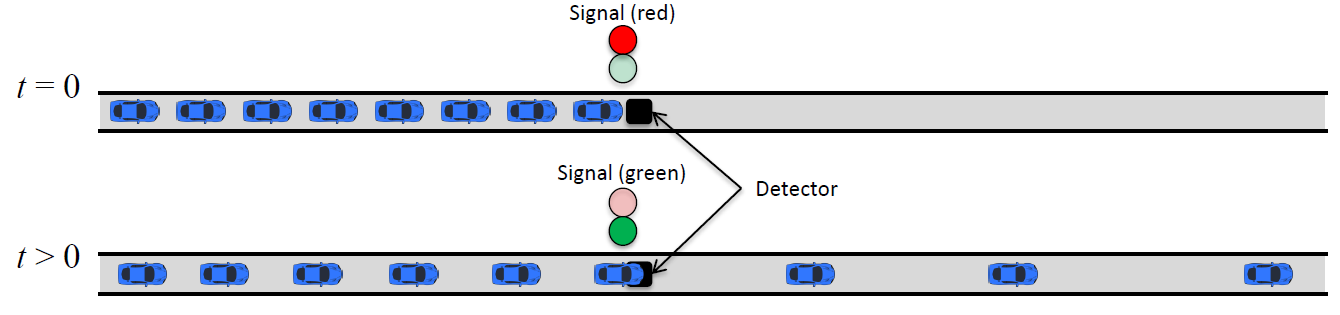}
\caption{Signal turns green at time $t=0$, and vehicles start
moving.
The first vehicle has free road ahead.}
\label{fig-basic-experiment}
\end{figure}

{\bf Experiment with a free road ahead.}
\newline
Consider a setup presented in Figure~\ref{fig-basic-experiment}.

\begin{figure}
\centering
\includegraphics[width=\textwidth]{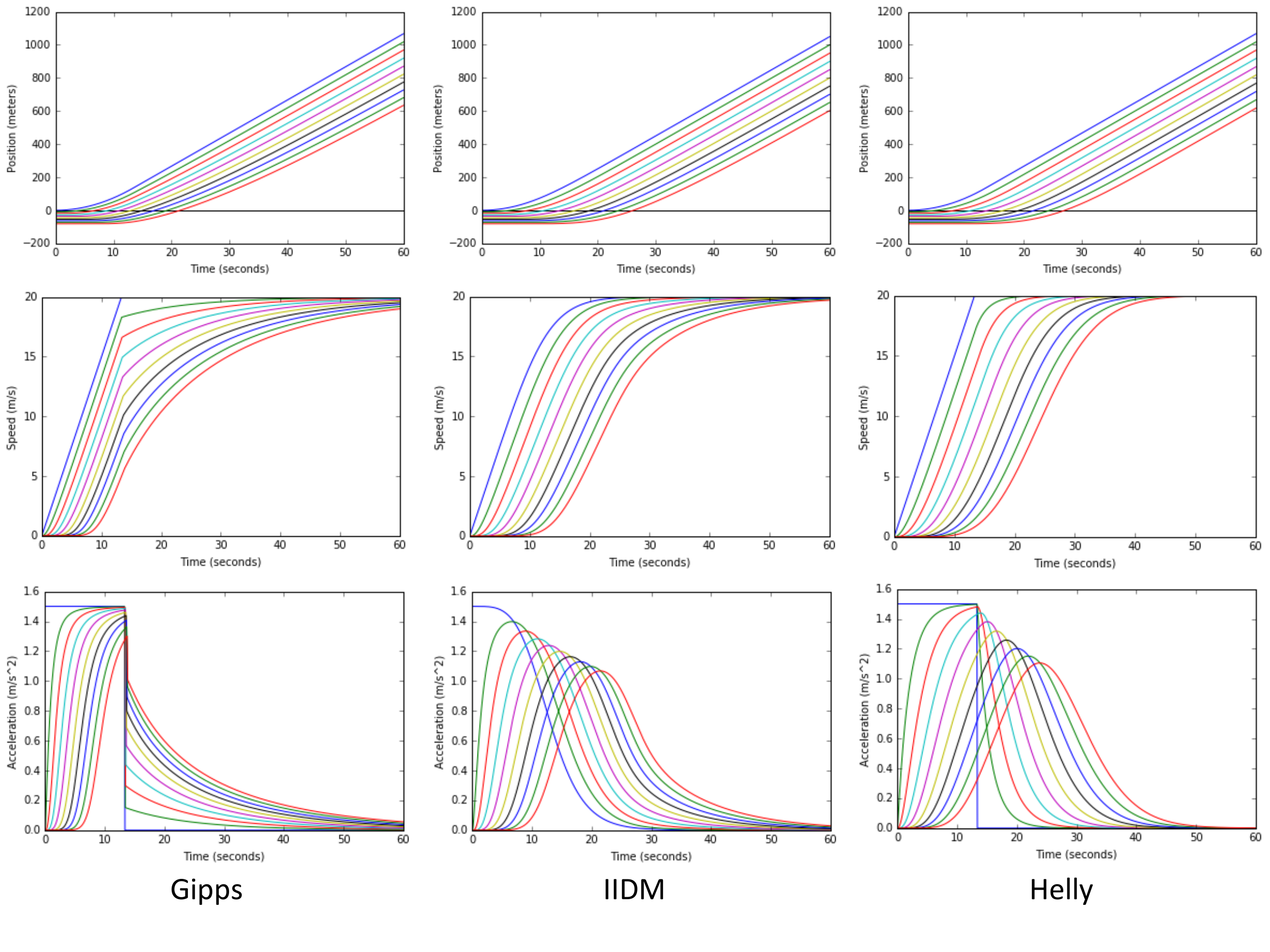}
\caption{Experiment with a free road ahead: comparison of vehicle
trajectories, speeds and accelerations between the three car following
models.}
\label{fig-trajectories-free}
\end{figure}

\begin{figure}[h]
\centering
\includegraphics[width=\textwidth]{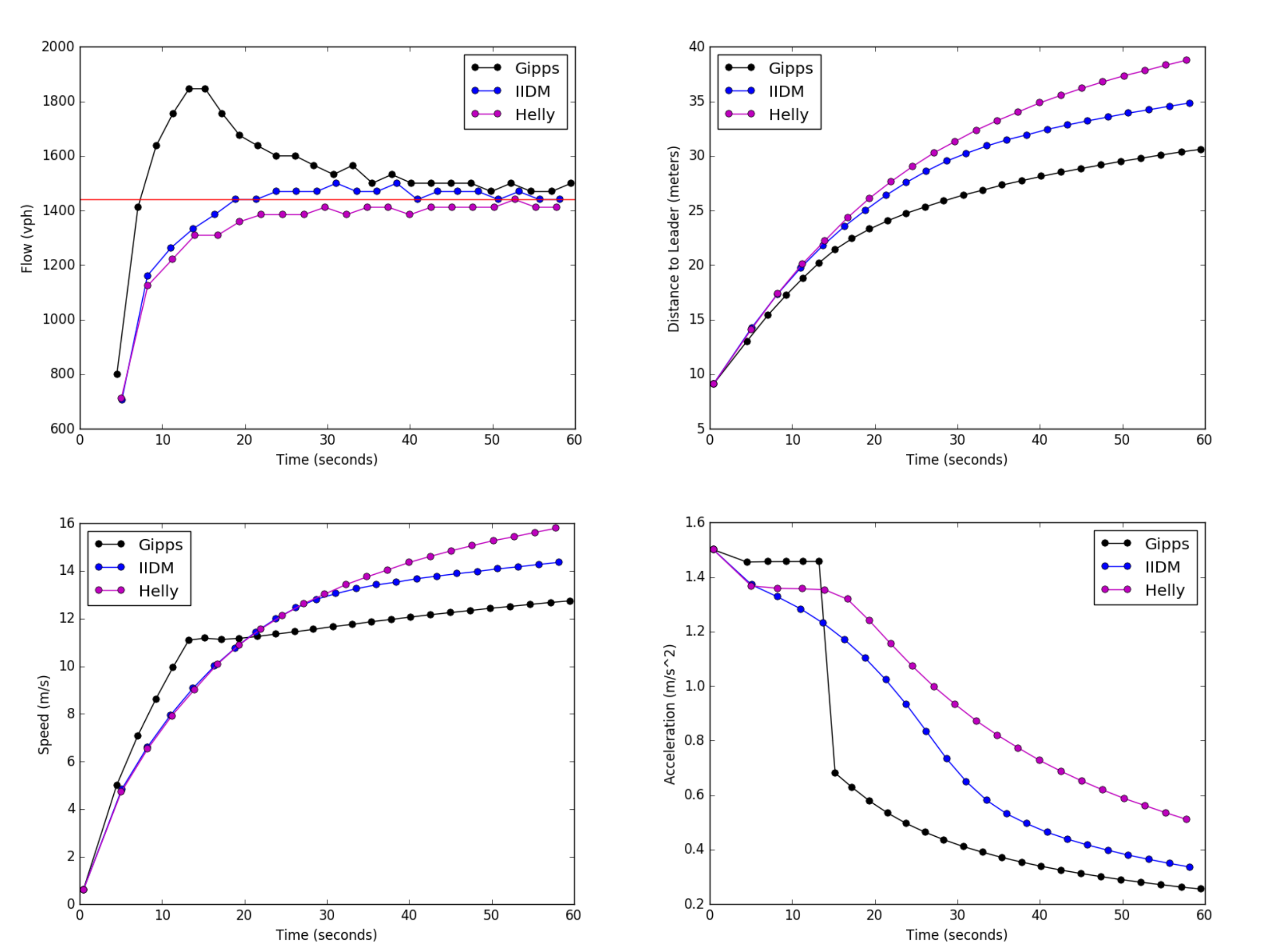}
\caption{Experiment with a free road ahead: comparison of point
measurements of flow, distance to leader, speed and acceleration
at the detector location between the three car following models.}
\label{fig-model-comparison-free}
\end{figure}

The initial condition at time $t=0$ is that infinite number of vehicles
are standing in the queue with the minimal gap between them.
The light turns green, and vehicles are released.

Figure~\ref{fig-trajectories-free} shows trajectories, speeds and
accelerations of the first ten vehicles from the queue computed
by the three car following models.
The signal is located at position 0 indicated by the horizontal black
line in the top three trajectory plots.
The first vehicle is governed by the car following model just
as everyone else, but its leader is infinitely far.
From the acceleration and speed plots one can see that in the Gipps
and the Helly models the first vehicle accelerates with maximal
acceleration $a_{\max}$ until reaching the maximal speed, at which point
the acceleration instantaneously drops to 0.
In the IIDM, the first vehicle accelerates with $a^\ast(t)$
from equation~\eqref{eq_iidm_afree}, approaching the maximal
speed asymptotically.

The most important for the intersection throughput assessment, however,
is the traffic behavior at the stop bar --- at the detector location
indicated in Figure~\ref{fig-basic-experiment}.
Figure~\ref{fig-model-comparison-free} presents the point measurments
obtained from this detector location, with each dot corresponding to
a vehicle passing the detector.
Flow (top left) is computed for a vehicle passing the detector based on the
time passed after the previously detected vehicle,
taken as a headway $\theta(t)$, which is then inverted ($f(t) = 1/\theta(t)$)
and converted to vehicles per hour (vph).
The red horizontal line corresponds to the equilibrium flow,
in our case --- 1440 vph, when vehicles move at maximal speed.
Gap (top right) between vehicles as well as speed (bottom left)
are monotonically increasing, while acceleration (bottom right)
is monotonically decreasing.

As is evident from plots in Figure~\ref{fig-model-comparison-free},
Gipps model produces rather aggressive
car following pattern to the point that it manages to push through
the intersection 26 vehicles per minute, two more than would pass through
the intersecton in an equilibrium flow (24 vehicles per minute),
while IIDM and Helly model push through 23 and 22 vehicles per minute
respectively --- see the middle row (free road ahead) of
Table~\ref{tab-model-throughput}.
What is interesting about this observation, is that with the Gipps model
one could argue that a signal would increase the road throughput by creating
pulses in the vehicle flow, as the one in
Figure~\ref{fig-model-comparison-free} (top left).
Moreover, the smaller the signal cycle, the bigger will be the throughput
increase.
This is counterintuitive and, likely, unrealistic.
IIDM and Helly model can be tuned to behave more agressively by increasing
their parameters $\delta_1,\delta_2$ in IIDM and $\alpha_1,\alpha_2$ in Helly.
Neither of these two models, however, can reach the throughput result of
Gipps.

\begin{figure}[h]
\centering
\includegraphics[width=\textwidth]{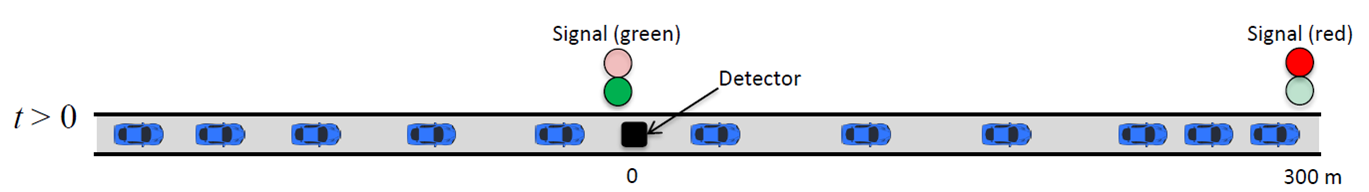}
\caption{Signal turns green at time $t=0$, and vehicles start
moving.
The first vehicle encounters red light in the next intersection 300 meters
downstream.}
\label{fig-experiment-with-stop}
\end{figure}

{\bf Experiment with a red light downstream.}
\newline
Let us modify the experiment setup by introducing the second intersection
downstream of the first one, where vehicles have to stop at the red
light --- Figure~\ref{fig-experiment-with-stop} depicts the modified
configuration.

\begin{figure}[h]
\centering
\includegraphics[width=\textwidth]{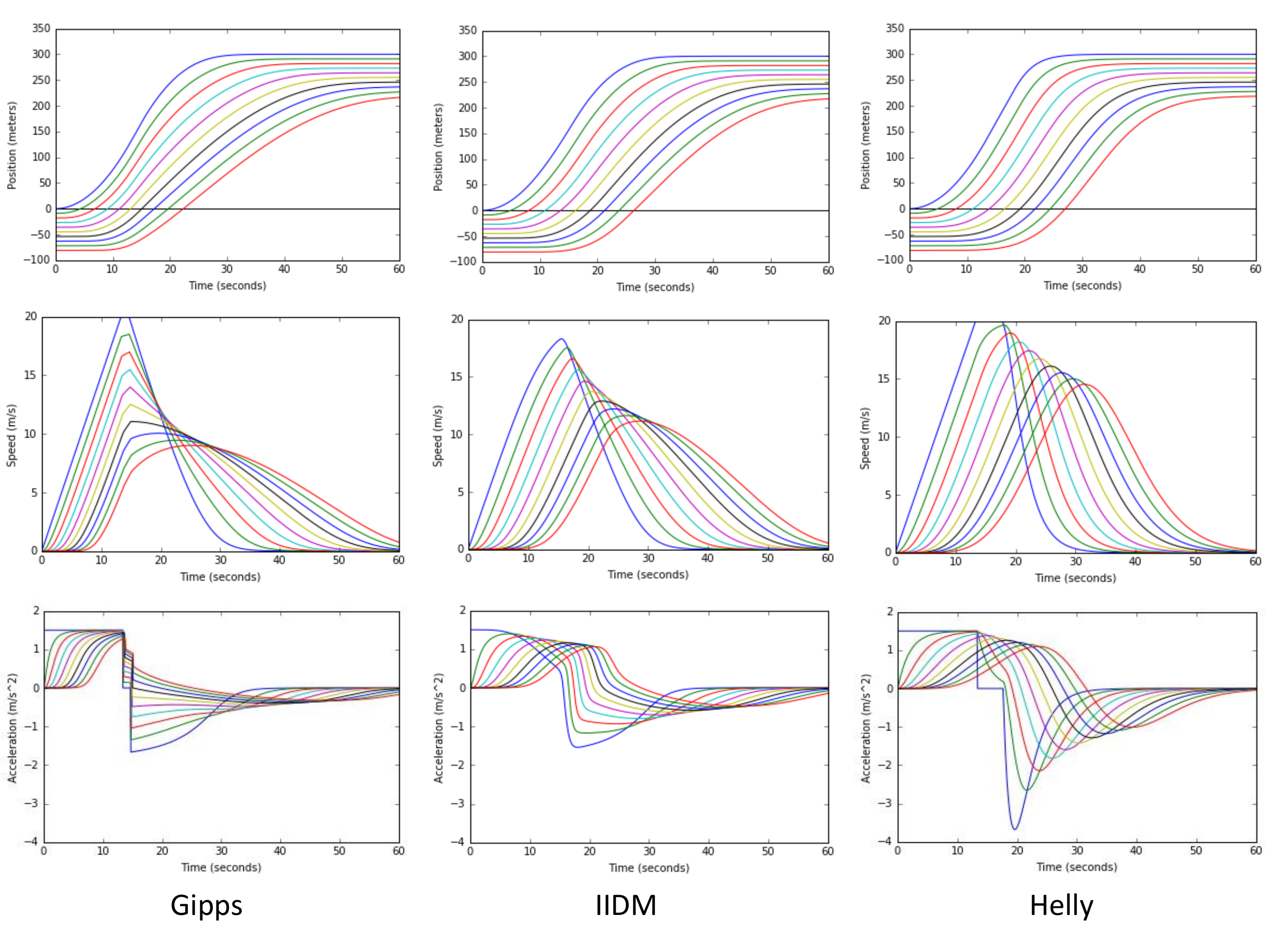}
\caption{Experiment with a red light ahead: comparison of vehicle trajectories,
speeds and accelerations between the three car following models.}
\label{fig-trajectories-stop}
\end{figure}

\begin{figure}[h]
\centering
\includegraphics[width=\textwidth]{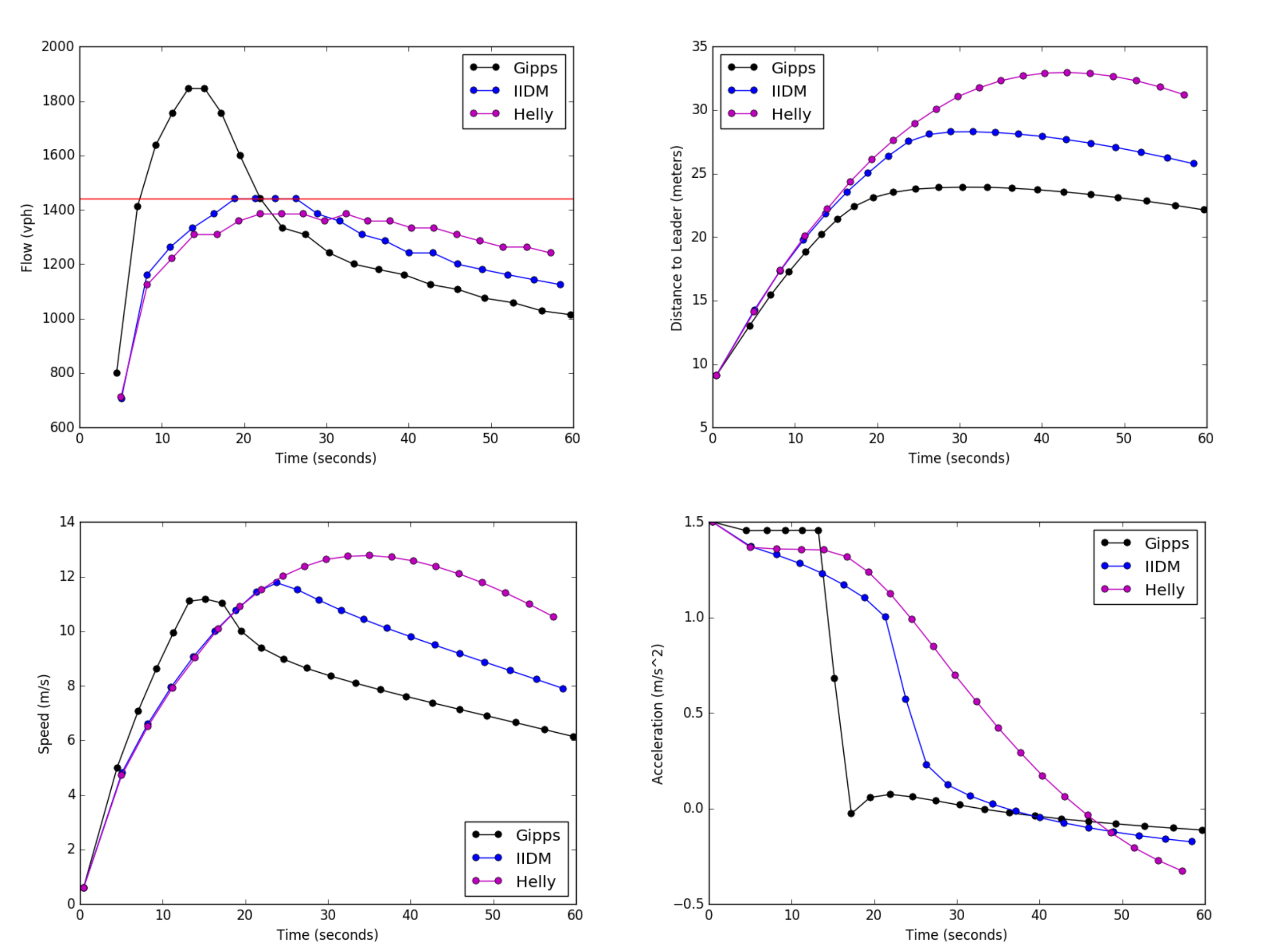}
\caption{Experiment with a red light ahead: comparison of point
measurements of flow, distance to leader, speed and acceleration
at the detector location between the three car following models.}
\label{fig-model-comparison-stop}
\end{figure}

In this experiment, the second intersection is 300 meters away from the first
one.
This distance is enough to hold 33 vehicles in the queue
(see Table~\ref{tab-micro-notation} for default values of the car length $l$
and the minimal gap $g_{\min}$), which is more than the most aggressive,
Gipps, car following model can send in one minute.

It is important to note that if instead of any car following model
we were using a Point Queue model with limited or unlimited queues,
such as in~\cite{liorisetal15}, we would be able to send as many
vehicles through the first intersection, as our saturation flow
setting would allow.
In the case of our example, if we set the saturation flow
to 24 vehicles per minute (equal to our equilibrium flow), after
1 minute of green light in the first intersection, 24 vehicles
would be transferred from the first queue to the next, in front of
the second intersection.
The car following model, on the other hand, exhibits the braking
effect that propagates back and reduces the vehicle flow though the
first intersection.
In the experiment with the red light downstream we study the impact of this
braking effect on the throughput of the first intersection.

Figure~\ref{fig-trajectories-stop} shows trajectories, speeds and
accelerations of the first ten vehicles from the queue computed
by the three car following models.
The first signal is at position 0 indicated by the horizontal black
line in the top three trajectory plots, 
and the second signal is at position $x_s=300$.
To make the first vehicle stop at $x_s$, we place a ``blocking vehicle''
in front at position
\[ x_b = x_s + g_{\min} + l = 300 + 4 + 5 = 309,\]
with velocity $v_b=0$.
Governed by the car following model, the first vehicle
stops at position $x_s$ to maintain the minimal gap
with this virtual ``blocking vehicle''.
In our case, the first vehicle in the Gipps and the Helly models reaches
the maximal speed before starting to brake.
Moreover, in the Helly model the first vehicle continues with the maximal
longer than in Gipps, allowing the second vehicle to almost
reach the maximal speed, which then leads to prohibitively
sharp deceleration jump.
In contrast, in IIDM the first vehicle starts to brake earlier
than in Gipps and Helly models, even before reaching the maximal speed,
resulting in smooth speed curves.

Point measurements taken at the detector location, shown in
Figure~\ref{fig-experiment-with-stop}, and presented in
Figure~\ref{fig-model-comparison-stop}, indicate the reduction
in flow through the first intersection as the result of the
braking propagation.
Gipps, IIDM and Helly models manage to send 22, 21 and 21 vehicles per minute
respectively through the first intersection --- see
the middle row (red light ahead) of
Table~\ref{tab-model-throughput}.
As before, the red horizontal line in the top left plot
corresponds to the equilibrium flow,
in our case --- 1440 vph, when vehicles move at maximal speed.
These plots also indicate the reactiveness of the studied car following models:
how fast the cars upstream react to the behavior of the first vehicle.
Given the IIDM and Helly model parameter values from
Table~\ref{tab-model-defaults}, Gipps model has the shortest reaction time,
followed by IIDM and then, by Helly model.\footnote{Reactiveness of
the two latter models can be somewhat increased by increasing parameters
$\delta_1, \delta_2$ for IIDM and $\alpha_1, \alpha_2$ for Helly model.}

{\bf Experiment with different accelerattion levels.}
\newline
Now, let us explore how throughput of the first intersection in our
two previous experiments depends on the maximal acceleration $a_{\max}$.
We repeat both experiments, with the free road and with the red light ahead,
for
three different values of $a_{\max}$: $0.8$, $1.5$ (our default) and 
$2.5 \mbox{ m/s}^2$.

\begin{figure}[h]
\centering
\includegraphics[width=\textwidth]{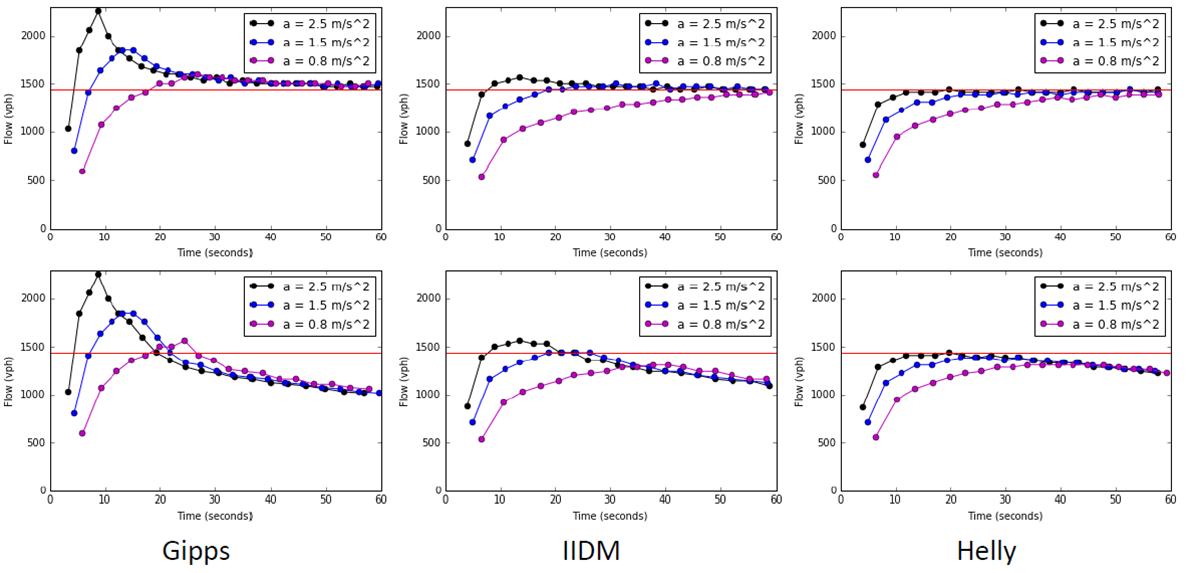}
\caption{Comparing flows for different values of $a_{\max}$
for two experiments --- with the free road (top), and
with the red light ahead (bottom).}
\label{fig-acceleration-comparison}
\end{figure}

\begin{table}[h]
\begin{center}
\begin{tabular}{|c|c|c|c|c|}
\hline
Value of $a_{\max}$ & Type of experiment & Gipps model & IIDM & Helly model \\
\hline
\multirow{2}{*}{$0.8\mbox{ m/s}^2$} &
free road ahead & 23 & 20 & 20 \\
 & red light ahead & 20 & 19 & 20 \\
\hline
\multirow{2}{*}{$1.5\mbox{ m/s}^2$} &
free road ahead & {\bf 26} & 23 & 22 \\
& red light ahead & 22 & 21 & 21 \\
\hline
\multirow{2}{*}{$2.5\mbox{ m/s}^2$} &
free road ahead & {\bf 27} & {\bf 24} & 23 \\
& red light ahead & 22 & 22 & 22\\
\hline
\end{tabular}
\end{center}
\caption{Summary of three experiments --- intersection
throughput in vehicles per minute.
Values equal to or exceeding the equilibrium flow, 24 vehicles per minute,
are given in bold.}
\label{tab-model-throughput}
\end{table}

Figure~\ref{fig-acceleration-comparison} presents the point measurements
obtained from the detector location for the cases of free road (top)
and red light downstream (bottom).
Table~\ref{tab-model-throughput} summarizes the throughput results
for all the model-accelertion-experiment combinations.

The main findings of this experiment are:
\begin{itemize}
\item with low $a_{\max}$ and braking effect, Helly model produces
larger throughput than IIDM, whereas generally the opposite is true;
\item with low $a_{\max}$, IIDM and Helly model behave similarly;
\item with high $a_{\max}$ and braking effect, the all three models
produce the same throughput;
\item braking effect reduces the impact of $a_{\max}$ parameter on throughput.
\end{itemize}

To perform a spacial analysis of the traffic flow shock wave propagation
for different values of $a_{\max}$,
we have to translate the car following behavior into
a \emph{macroscopic} model, which we do next.

\section{Micro-to-Macro Translation}\label{sec_micro_macro}
Macroscopic models describe traffic in terms of density, and speed.
The road is divided into $1, \dots, N$ links, and the state of
the system at time $t$ is given by the density-speed pair
$\{\rho_i(t), V_i(t)\}_1^N$.
Table~\ref{tab-macro-notation} contains the notation used in macro-modeling. 
\begin{table}[h]
\begin{center}
\begin{tabular}{|l|l|l|}
\hline
Symbol & Description & Default value\\
\hline
$\Delta x_i$ & Length of link $i$. & $\Delta x_i = 5$ m.\\
$t$, $\Delta t$ & Time and the model time step
(same as in the car following model). & $\Delta t = 0.05$ s.\\
$\rho_i(t)$ & Vehicle density in link $i$. & \\
$\rho_J$ & Maximal admissible (jam) density. & $\rho_J=1/9$ veh. per meter.\\
$V_i(t)$ & Average traffic speed link $i$. &\\
$v_{\max}$ & Maximal admissible traffic speed
(same as in the car following model). & $v_{\max}=20$ m/s.\\
$f_i(t)$ & Vehicle flow out of link $i$. &\\
$f_0(t)$ & Vehicle flow entering link $1$. &\\
\hline
\end{tabular}
\end{center}
\caption{Macro-modeling notation.}
\label{tab-macro-notation}
\end{table}

Vehicle density $\rho_it)$ is computed from the gap between vehicles and the
vehicle length:
\be
\rho_i(t) = \frac{1}{g_i(t) + l},
\label{eq_mmr_rho}
\ee
where notation $g_i(t)$ defines average gap between vehicles that are
in link $i$ at time $t$.
Obviously, $\rho_i(t)\leq\rho_J = \frac{1}{g_{\min}+l}$.

Every time step, density values in each link are updated according
to the conservation law:
\be
\rho(t+\Delta t) = \rho_i(t) + \frac{\Delta t}{\Delta x_i}\left(f_{i-1}(t) - f_i(t)\right), \;\;\;
i = 1,\dots,N,
\label{eq_macro_rho}
\ee
where $f_i(t) = \rho_i(t)V_i(t)$,
and the entering flow $f_0(t)$ is given.

The speed equation is derived from vehicle speed.
For points $(t, x)$ on the trajectory of a vehicle,
we can write:
\[ v(t) = V(t, x) .\]
The change in position during one time step $\Delta t$ can be expressed
by the first order Taylor expansion around $(t, x)$:
\begin{eqnarray*}
v(t+\Delta t) & = & V(t+\Delta t, x+v(t)\Delta t) \\
& = & V(t, x) + \frac{\partial V(t, x)}{\partial t}\Delta t +
\frac{\partial V(t, x)}{\partial x}v(t)\Delta t \\
& = & V(t, x) + \left(\frac{\partial V}{\partial t} + V(t,x)\frac{\partial V}{\partial x}\right)\Delta t.
\end{eqnarray*}
At the same time, from the car following model we have:
\[ v(t + \Delta t) = v(t) + a(t)\Delta t .\]
Thus, we get:
\[ V(t, x) + \left(\frac{\partial V}{\partial t} +
V(t,x)\frac{\partial V}{\partial x}\right)\Delta t =
V(t, x) + a(t)\Delta t, \]
which, after cancelling $V(t, x)$ and dividing both sides of the equation
by $\Delta t$, yields:
\[\frac{\partial V}{\partial t} + V(t,x)\frac{\partial V}{\partial x} = a(t).\]
Discretizing this equation in time and space, we can write:
\[ \frac{V_i(t+\Delta) - V_i(t)}{\Delta t} +
V_i(t) \frac{V_i(t)-V_{i-1}(t)}{\Delta x_i} = a_i(t) , \;\;\; i=1,\dots,N, \]
and $V_0(t) = V_1(t)$.

Thus, we obtain the speed equation:
\be
V_i(t+\Delta t) = V(t) +
\left(a_i(t) - V_i(t)\frac{V_i(t)-V_{i-1}(t)}{\Delta x_i}\right)\Delta t.
\label{eq_macro_speed}
\ee

Here $a_i(t)$ is defined by the car following model.
\begin{itemize}
\item For Gipps model, following~\eqref{eq_gipps}, we have:
\be
a_i(t) = \min\left\{a_{\max}, \frac{v_{\max}-V_i(t)}{\Delta t}, 
\frac{-V_i(t) - b\tau
+ \sqrt{(b\tau)^2 + (V_{i+1}(t))^2 +
2b\left(\frac{1}{\rho_i(t)}-\frac{1}{\rho_J}\right)}}{\Delta t}\right\},
\label{eq_macro_gipps}
\ee
$i=1,\dots,N$; $V_{N+1}=v_{\max}$.
\item For IIDM, following~\eqref{eq_iidm}, we have:
\be
a_i(t) = \left\{\begin{array}{ll}
a_{\max}\left(1 - \frac{g_{d,i}(t)}{g_i(t)}\right), &
\mbox{ if } \frac{g_{d,i}(t)}{g_i(t)} \geq 1,\\
a_i^\ast(t)\left(1 - \left(\frac{g_{d,i}(t)}{g_i(t)}
\right)^{\delta_1 a_{\max}/a_i^\ast(t)}
\right), & \mbox{ otherwise},
\end{array}\right. 
\label{eq_macro_iidm0}
\ee
where
\be
a_i^\ast(t) =
a_{\max}\left(1-\left(\frac{V_i(t)}{v_{\max}}\right)^{\delta_2}\right),
\label{eq_macro_iidm_afree}
\ee
while the actual gap, $g_i(t)$, and the desired gap,
$g_{d,i}(t)$, can be expressed through density:
\be
g_i(t) = \frac{1}{\rho_i(t)}-l, \;\;\; i=1,\dots,N,
\label{eq_macro_iidm_gap}
\ee
and
\be
g_{d,i}(t) = \frac{1}{\rho_J}-l + \max\left\{0,
V_i(t)\tau + \frac{V_i(t)(V_i(t)-V_{i+1}(t))}{2\sqrt{a_{\max}b}}\right\}, \;\;
i=1,\dots,N; \;\; V_{N+1}=v_{\max}. 
\label{eq_macro_iidm_desired_gap}
\ee
\item For Helly model, following~\eqref{eq_helly}, we have:
\be
a_i(t) = \min\left\{a_{\max}, \frac{v_{\max}-V_i(t)}{\Delta t}, 
\alpha_1\left(V_{i+1}(t)-V_i(t)\right) + 
\alpha_2\left(\frac{1}{\rho_i(t)}-\frac{1}{\rho_J}-V_i(t)\tau\right)\right\},
\label{eq_macro_helly}
\ee
$i=1,\dots,N$; $V_{N+1}=v_{\max}$.
\end{itemize}

\begin{figure}[h]
\centering
\includegraphics[width=\textwidth]{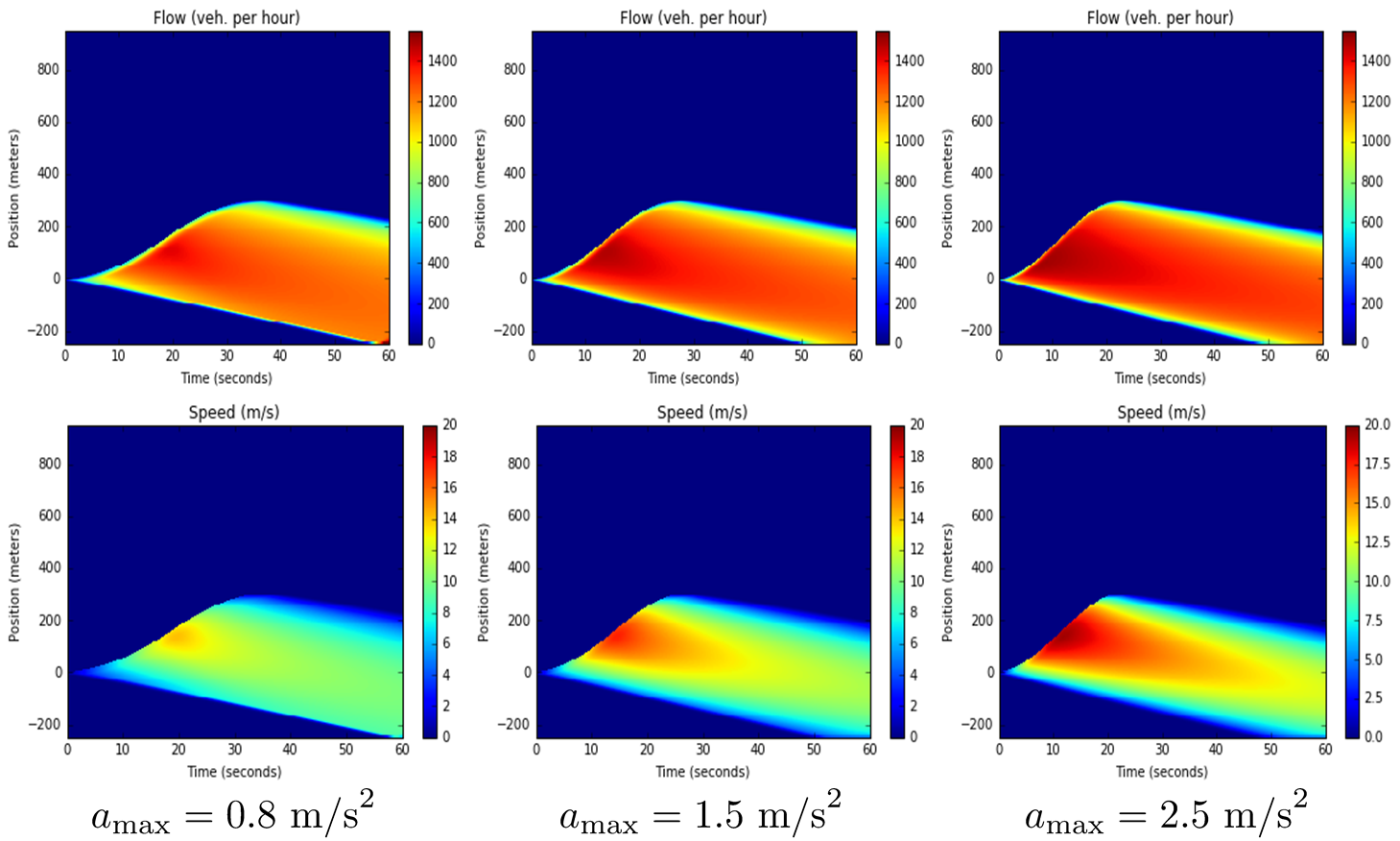}
\caption{Flow (top) and speed (bottom) contours generated by the
IIDM-induced macroscopic
model in the experiment with the red light downstream for
$a_{\max} = 0.8$, $1.5$ and $2.5\mbox{ m/s}^2$.}
\label{fig-macro-contours}
\end{figure}

As an example,
we reproduced the experiment with the red light ahead and the red light
at the second intersection in the macroscopic environment.
The road is split into 240 links, each with length $l=5$ meters.
The first signal is located at link 50.
The initial condition is:
\begin{eqnarray*}
& & \rho_i(0) = \rho_J = \frac{1}{g_{\min}+l} = \frac{1}{4+5} =1/9, \;\;\; 1=1,\dots,49; \\
& & \rho_i(0) = 0, \;\;\; i = 50,\dots,N=240; \\
& & V_i(0) = 0, \;\;\; i = 1,\dots,N=240.
\end{eqnarray*}
The second signal with the red light is in link 110 (300 meters downstream of
the first one), which translates into condition:
\[ f_{110}(\cdot) = 0. \]

Figure~\ref{fig-macro-contours} presents the flow and speed contours,
produced by the simulation of 60 seconds of the IIDM-induced macroscopic model
for $a_{\max}=0.8, 1.5$ and $2.5\mbox{ m/s}^2$.
Here, the horizontal axis represents time in seconds and
the vertical axis --- space in meters, where cars travel from bottom to top.
The locations of the first and the second signals are at positions 0 and 300
on the vertical axis respectively.

\section{Effect off ACC and CACC}\label{sec_acc_cacc}
We will now explore the impact of ACC and CACC vehicles on intersection
throughput.
To do that, we repeat two experiments described in
Section~\ref{sec_cfm} --- the case of free road
and the case of red light downstream --- but this time,
throwing ACC and CACC vehicles into the traffic mix.
Values of car following parameters for ordinary, ACC- and CACC-enabled vehicles
are given in Table~\ref{tab-params-per-vehicle-type}.
As we can see, ACC and CACC vehicles can maintain shorter distances
to the car in front.

\begin{table}[h]
\begin{center}
\begin{tabular}{|l|c|c|}
\hline
Vehicle type & Reaction time $\tau$ (seconds) & Minimal gap $g_{\min}$ (meters)\\
\hline
Ordinary & 2.05 & 4 \\
ACC-enabled & 1.1 & 3 \\
CACC-enabled & 0.8 & 3 \\
\hline
\end{tabular}
\end{center}
\caption{Values of reaction time $\tau$ and minimal gap $g_{\min}$
for ordinary, ACC- and CACC-enabled vehicles.}
\label{tab-params-per-vehicle-type}
\end{table}

\begin{figure}[h]
\centering
\includegraphics[width=\textwidth]{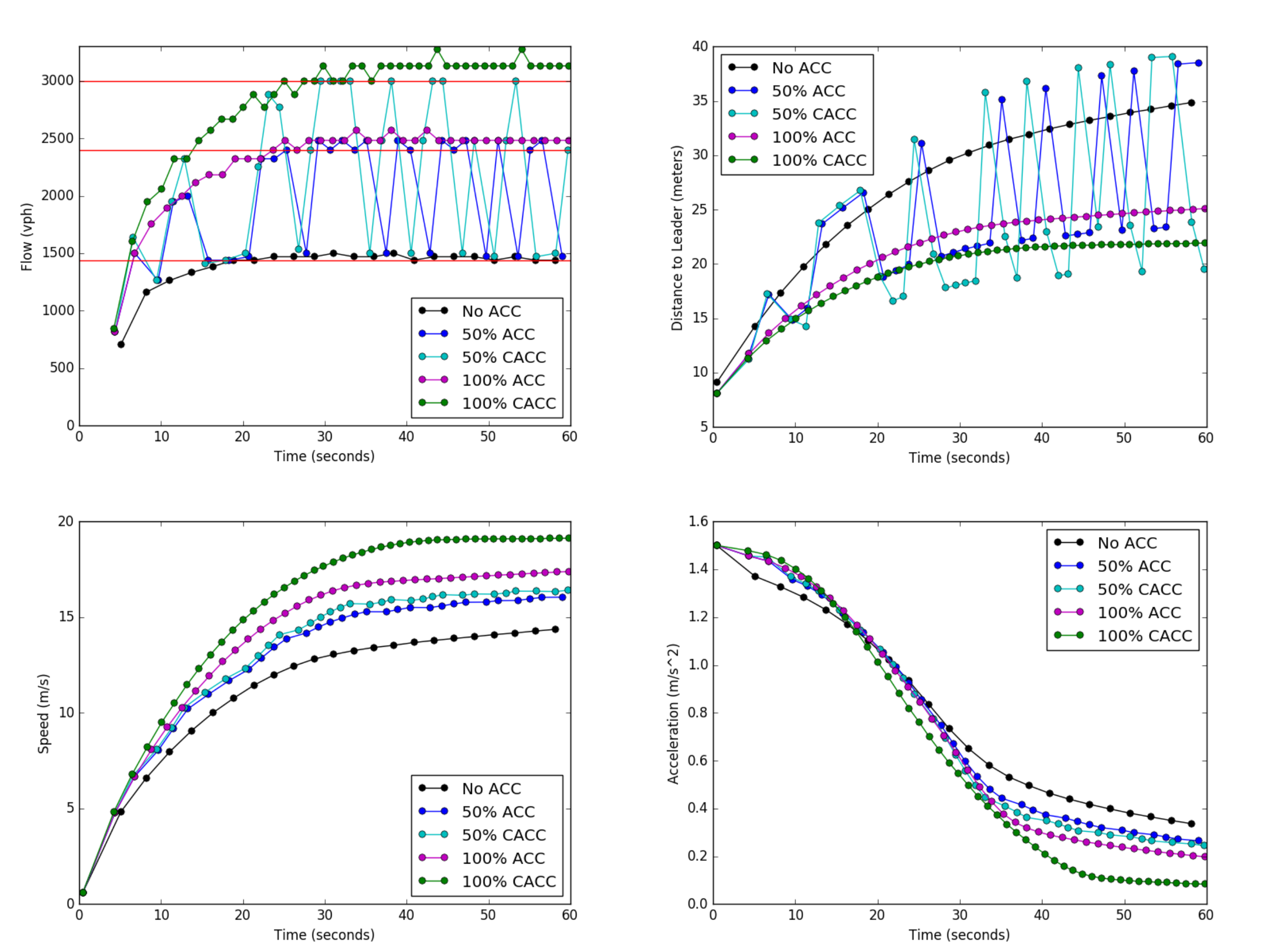}
\caption{Single intersection case: comparison of point measurements of flow,
distance to leader, speed and acceleration at the detector location
for different portions of ACC/CACC traffic.}
\label{fig-acc-portion-free}
\end{figure}

\begin{figure}[h]
\centering
\includegraphics[width=\textwidth]{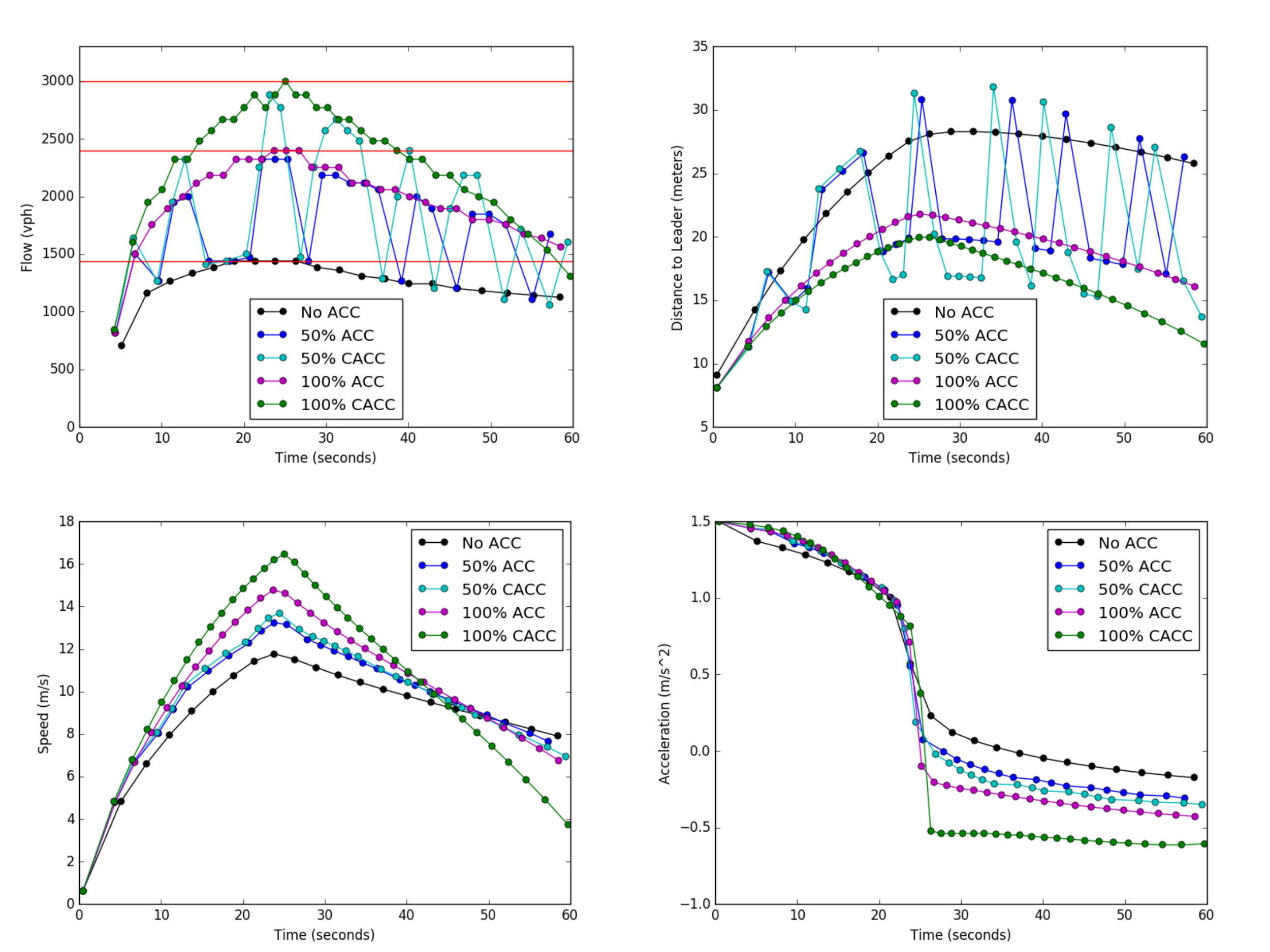}
\caption{Case with red light downstream: comparison of point measurements of flow,
distance to leader, speed and acceleration at the detector location
for different portions of ACC/CACC traffic.}
\label{fig-acc-portion-stop}
\end{figure}

\begin{figure}[h]
\centering
\includegraphics[width=\textwidth]{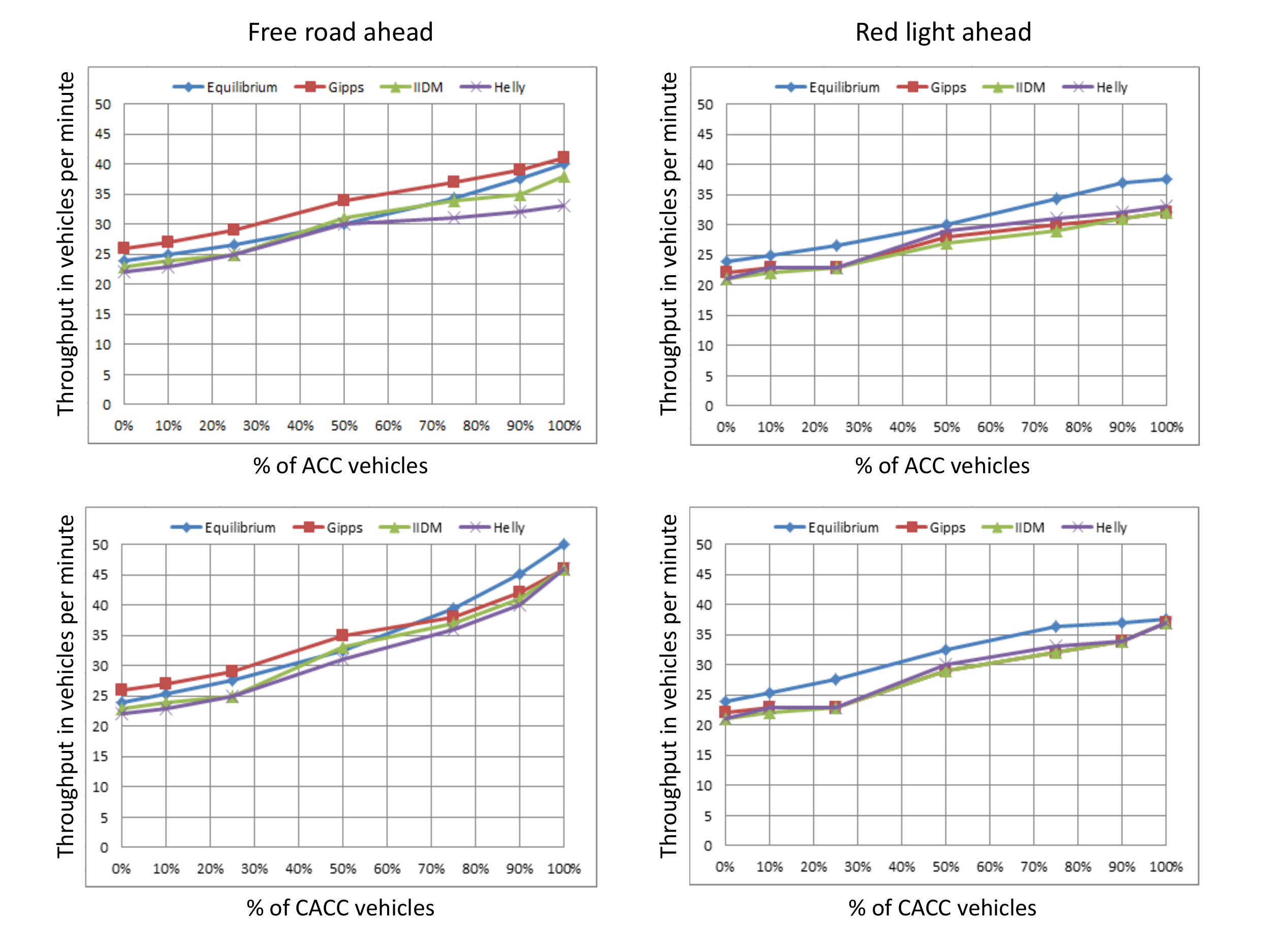}
\caption{Intersection throughput as a function ACC/CACC portion of traffic
computed with Gipps, IIDM and Helly car following models for two
cases --- free road downstream (left) and red light downstream (right).}
\label{fig-acc-cacc-per-model}
\end{figure}

We assume that the ACC vehicle has the same car following model as the
ordinary one, just with different $\tau$ and $g_{\min}$.
CACC vehicle behaves just as ACC, with ACC $\tau$ and $g_{\min}$, if
it follows an ordinary vehicle, but if it has another CACC car in front,
it assumes different car following behavior, which we call
\emph{CACC car following model}.

Denote $a_{IIDM}(t)$ the acceleration function defined by~\eqref{eq_iidm}.
Define \emph{constant-acceleration heuristic (CAH)} acceleration 
function~\cite{treiberkesting_book} (Chapter 11):
\be
a_{CAH}(t) = \left\{\begin{array}{ll}
\frac{v^2(t)\bar{a}_l(t)}{v_l^2(t)-2(x_l(t)-x(t)-l)\bar{a}_l(t)}, &
\mbox{ if } v_l(t)(v(t) - v_l(t)) \leq -2(x_l(t)-x(t)-l)\bar{a}_l(t),\\
\bar{a}_l(t) - \frac{(v(t)-v_l(t))^2\Theta(v(t)-v_l(t))}{2(x_l(t)-x(t)-l)} &
\mbox{ otherwise,}
\end{array}\right.
\label{eq_a_cah}
\ee
where
\[ \bar{a}_l(t) = \min\left\{\dot{v}_l(t), a_{\max}\right\}, \]
and
\[
\Theta(z) = \left\{\begin{array}{ll}
1, & \mbox{ if } z \geq 0,\\
0, & \mbox{ otherwise}.
\end{array}\right.
\]
Now we specify the CACC car following
model~\cite{treiberkesting_book} (Chapter 11):
\be
a_{CACC}(t) = \left\{\begin{array}{ll}
a_{IIDM}(t), & \mbox{ if } a_{CAH}(t) \leq a_{IIDM}(t), \\
a_{CAH}(t) + b\tanh\left(\frac{a_{IIDM}(t)-a_{CAH}(t)}{b}\right), &
\mbox{ otherwise.}
\end{array}\right.
\label{eq_cacc_model}
\ee

As before, we run the free road and the red light downstream experiments
using Gipps, IIDM and Helly car following models.
For each of these models, we compute the intersection throughput,
when portion of ACC (CACC) vehicles in the initial queue.
Thus, we evaluate 72 cases, each defined by:
(1) experiment --- free road or red light downstream;
(2) car following model --- Gipps, IIDM, Helly; 
(3) ACC or CACC; and (4) percentage of ACC (CACC) --- 10,
25, 50, 75, 90 and 100\%.

Figures~\ref{fig-acc-portion-free} and~\ref{fig-acc-portion-stop} compare
flows, gaps, speeds and acceleration obtained at the detector location
(see Figures~\ref{fig-basic-experiment} and~\ref{fig-experiment-with-stop})
for 0, 50 and 100\% ACC (CACC) penetration rate.
Three horizontal red lines on flow plots in both figures correspond
to equilibrium flows with 0\% ACC (CACC), with 100\% ACC and with
100\% CACC penetration rate.
These flows are computed as $3600/\theta_e$, where $\theta_e$ is given
by~\eqref{eq_min_headway} with $\tau$ and $g_{\min}$ from
Table~\ref{tab-params-per-vehicle-type}, yilding 1440, 2400 and 3000
vehicles per hour respectively.
In the flow and distance to leader plots, one can see how 50\% ACC curves
jump between the no ACC and 100\% ACC curves --- for ordinary vehicles
it is similar to the no ACC curve, and for an ACC vehicle it is
similar to 100\% ACC curve.
50\% CACC curves in the same plots jump between three curves --- no ACC,
100\% ACC and 10\% CACC.
This is because a CACC vehicle following an ordinary one behaves
like ACC vehicle. 

For a given ACC (CACC) penetration rate less than 100\%, the intersection
throughput is sensitive to the distribution of ACC (CACC) vehicles
in the initial queue.
For example, if 25\% all vehicles in the initial queue are ACC-enabled,
and all of them are concentrated at the head of the queue, we would get
higher vehicle count at the detector location after one minute, than
we would with 50\% ACC penetration rate when all ACC-enabled vehicles are
concentrated at the tail of the queue.
In another example, 50\% CACC penetration rate would not produce any gain
over 50\% ACC penetration rate, if ordinary and ACC/CACC vehicles
interleave --- one ordinary, one ACC/CACC, one ordinary, and so on --- since
CACC provides benefit over ACC in terms of throughput only when some
CACC vehicles have other CACC vehicles directly in front.

To mitigate this ACC (CACC) distribution bias, for each of the cases with
ACC (CACC) penetration rate less than 100\%, we run 100 one-minute simulations
of the three car following models and record vehicle counts at the
detector location, then take the median vehicle count.
For 100\% penetration rate the ACC (CACC) distribution is trivial, and hence,
a single simulation for each case is enough.
The intersection throughput results for all the 72 cases,
together with throughput values from Table~\ref{tab-model-throughput} obtained
for 0 ACC (CACC) penetration rate, are presented at the four plots in
Figure~\ref{fig-acc-cacc-per-model}.

Note that in each of the four plots in Figure~\ref{fig-acc-cacc-per-model},
in addition to the three curves corresponding to car following models,
there is a curve corresponding to the \emph{equilibrium} traffic flow.
These equilibrium curves are computed as follows.
Denote $\lambda\in[0,1]$ a portion of ACC (CACC) vehicles in the initial queue;
$\tau^{[C]ACC}$ and $g_{\min}^{[C]ACC}$ the reaction time and minimal
gap for ACC (CACC) vehicles, whose values are given in
Table~\ref{tab-params-per-vehicle-type}.
The average headway in the equilibrium state is obtained by modifying
expression~\eqref{eq_min_headway}:
\be
\theta(\lambda) = \lambda \tau^{[C]ACC} + (1-\lambda)\tau +
\frac{\lambda g_{\min}^{[C]ACC} + (1-\lambda)g_{\min} + l}{v_{\max}}.
\label{eq_modified_headway}
\ee
Then, the equilibrium flow in vehicles per minute is given by:
\be
f(\lambda) = 60/\theta(\lambda).
\label{eq_modified_max_flow}
\ee
This formula is sufficient for the case when there is a free road ahead.
In the case of the red light downstream, however, we are restricted
by the capacity of the link connecting the two intersections.
To account for that, we modify~\eqref{eq_modified_max_flow} accordingly:
\be
f(\lambda) = \min\left\{\frac{60}{\theta(\lambda)},
\frac{k\Delta}{\lambda g_{\min}^{[C]ACC} + (1-\lambda)g_{\min}+l}\right\},
\label{eq_modified_max_flow2}
\ee
where $\Delta$ is the length of the link between the two intersections,
and $k$ is the number of lanes in that link.
In our experiment, $\Delta = 300$, and $k=1$.

\section{Platoon Model}\label{sec_platoon_model}
\begin{figure}[h]
\centering
\includegraphics[width=\textwidth]{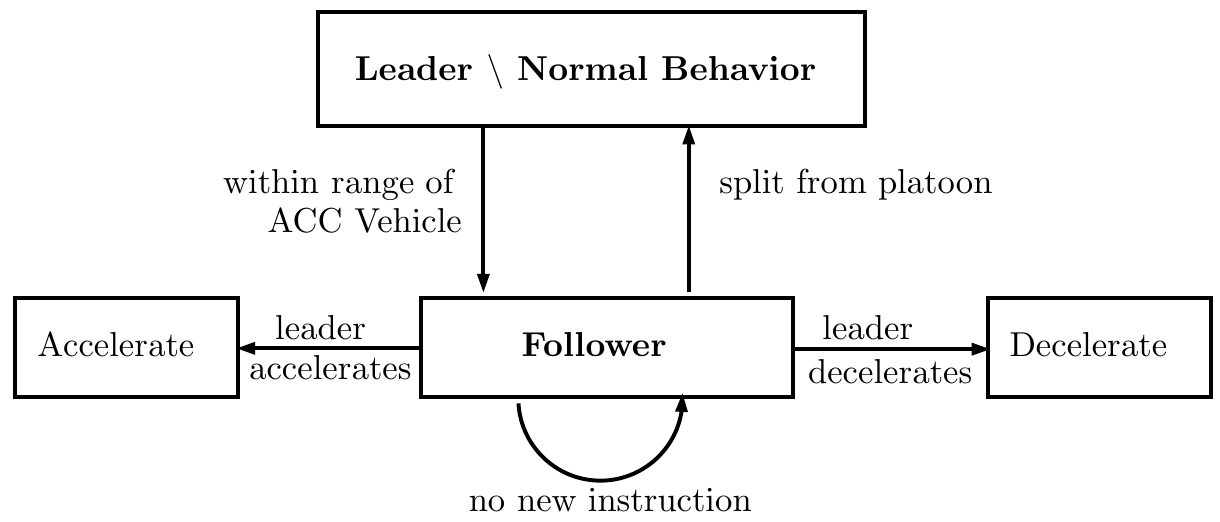}
\caption{State machine describing platoon behavior.}
\label{fig-state-machine}
\end{figure}

Vehicles equipped with CACC technology can communicate with one another to form platoons. These platoons can increase the throughput of intersections by decreasing headways between successive vehicles. In simulation, platoon management and formation is divided into three phases:
\begin{enumerate}
\item identifying vehicles that can be grouped into platoons;
\item Adjusting parameters of leaders and followers in platoons; and
\item performing maintenance on the platoon.
\end{enumerate}
This hierarchy is modeled by the state machine in Figure~\ref{fig-state-machine}.

\begin{figure}[h]
\centering
\includegraphics[width=\textwidth]{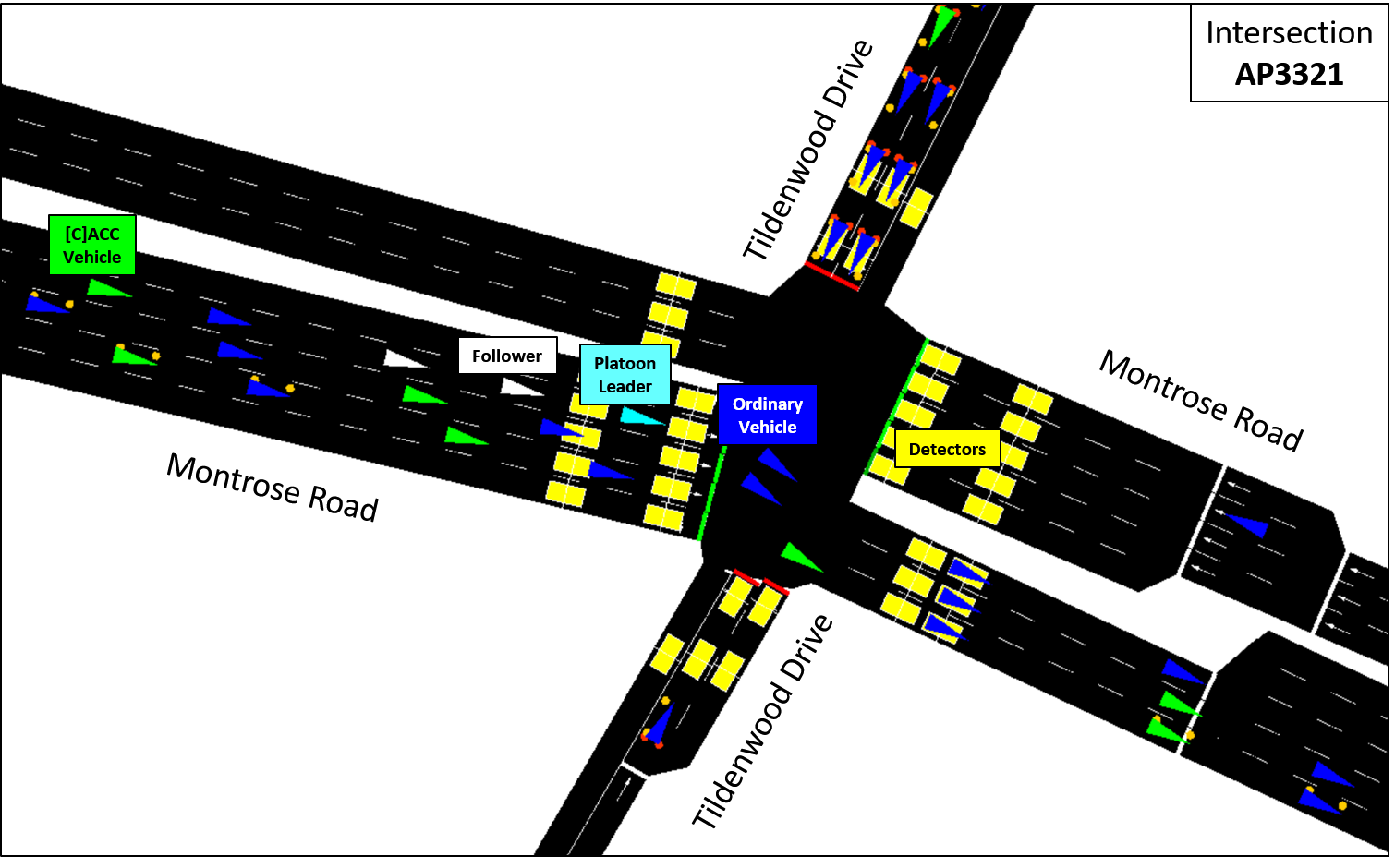}
\caption{SUMO screenshot --- ordinary, standalone [C]ACC vehicles and platoons are crossing
intersection AP3321 (Montrose Road and Tildenwood Drive).
Platoon leader and follower, as well as standalone [C]ACC and ordinary vehicles are labeled
accordingly.}
\label{fig-screenshot-ap3321}
\end{figure}

To form a platoon, vehicles must be in sequence
with one another on a given lane.
However, vehicles need not share the same final destination
and are free to switch lanes or leave the platoon if necessary.
If an intermediate vehicle in the platoon changes its route by
making a turn or changing lanes, the platoon splits into two:
one platoon for the vehicles ahead of the intermediate vehicle
and another for all the vehicles behind.

A platoon's lead vehicle has the same properties as ACC vehicles.
An isolated CACC vehicle is a leader of a platoon of size 1.
When a platoon leader comes into range of another CACC vehicle in front,
it joins the platoon becoming a follower.
Followers have reduced headway and travel much closer to one another
than standalone vehicles.
In addition, followers are able to receive information from the leader,
such as to accelerate after a green light at an intersection or to
decelerate approaching an obstacle, e.g. red light downstream.

Since followers are not bound to the same route as the platoon leader,
they are free to separate.
After leaving the platoon, the headway and acceleration parameters are
restored to their original values.
This can happen for example when the follower changes its route or
becomes separated from the rest of platoon, e.g., due to switching
traffic signal as it crosses the intersection.

Figure~\ref{fig-screenshot-ap3321} displays a screenshot of SUMO simulation run in graphical mode.
Ordinary vehicles are colored in blue.
[C]ACC vehicles with no followers (standalone) are colored in green.
Platoon leaders are colored in cyan, and followers are white.

Next, we discuss SUMO simulation results.

\section{Case Study: Simulation of North Bethesda Road Network}\label{sec_simulation}
\begin{figure}[h]
\centering
\includegraphics[width=\textwidth]{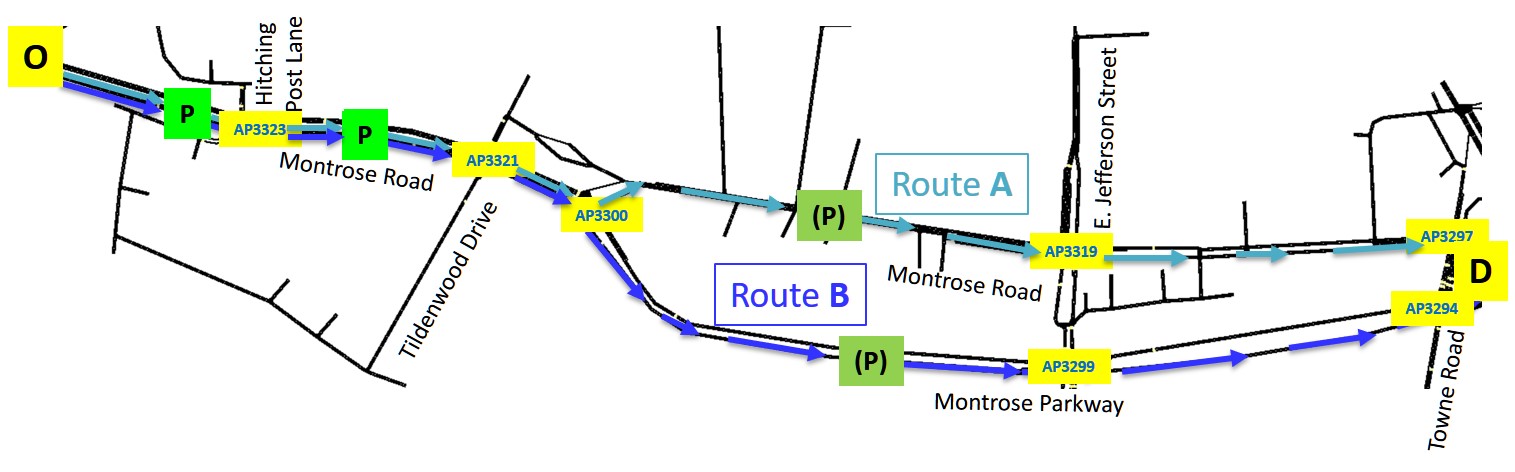}
\caption{North Bethesda network --- O-D pair with routes A and B.
Labels `P' indicate links, where platoons form, when platooning is enabled.
Labels `(P)' indicate additional links with optional platooning, when it is enabled.}
\label{fig-routes}
\end{figure}

To study the impact of platooning, we built the SUMO~\cite{sumo} simulation model
of the North Bethesda network with seven
major intersections, shown in Figure~\ref{fig-routes}.
IIDM and CACC car following models were implemented in C++ within SUMO, and
platoon management, presented as a state machine in Figure~\ref{fig-state-machine},
was implemented in Python using SUMO/TraCI API~\cite{sumo-traci}.
The corresponding source code repository can be accessed at~\cite{sumo-project}.

Using vehicle counts and estimated turn ratios, we generated
1 hour of origin-destination (O-D) travel demand data with a route
assigned to each vehicle.
These travel demand data together with signal plans constitute input for the simulation model.

We focused our study on the single O-D pair with two routes, A and B, connecting
origin (O) and destination (D) --- see Figure~\ref{fig-routes}.
Routes A and B coincide in the beginning, following Montrose Road, and split
at intersection AP3300 into Montrose Road (route A) and Montrose Parkway (route B).

\begin{figure}[h!]
\centering
\includegraphics[width=6in]{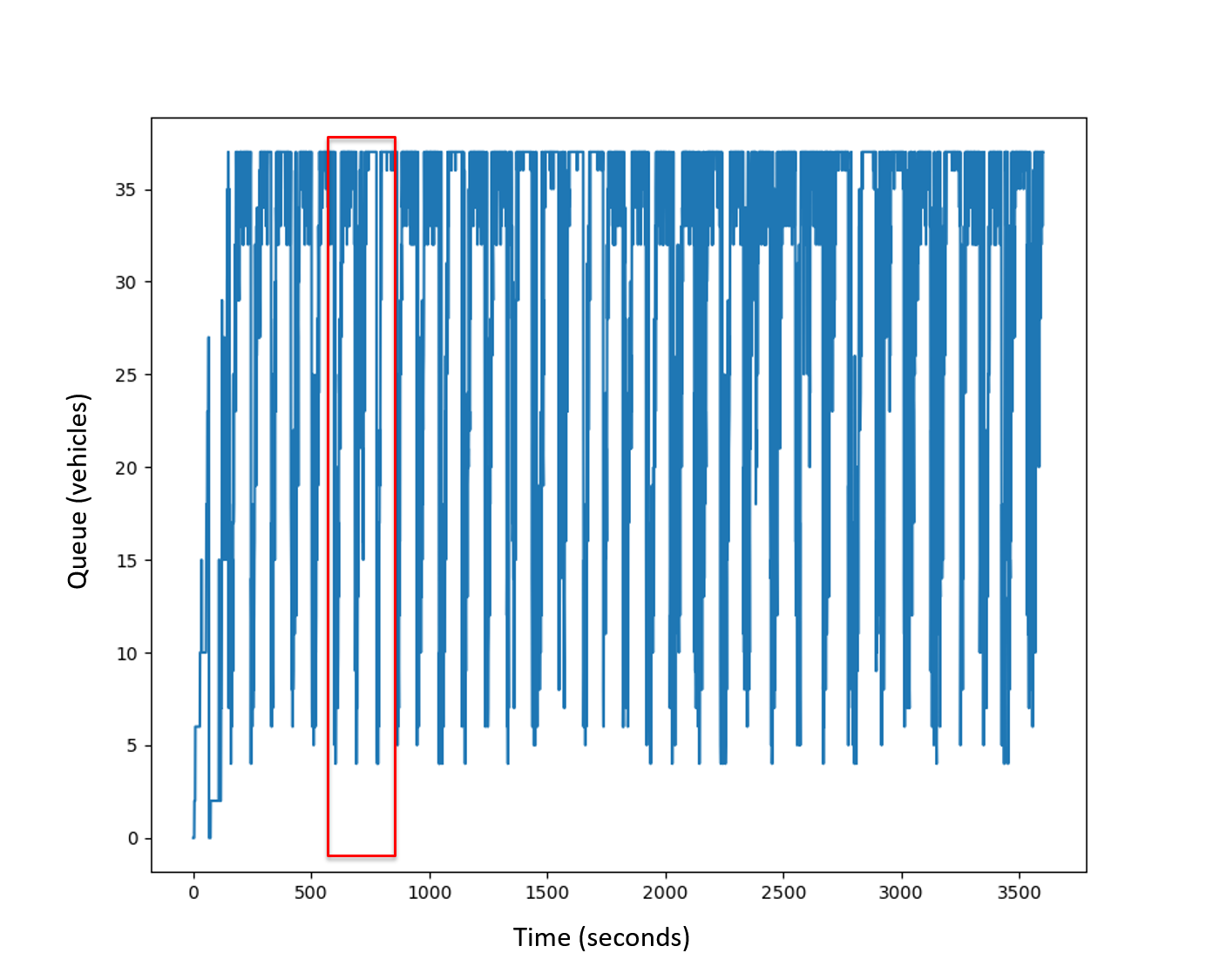}
\caption{Queue dynamics at the route A (B) approach to the intersection
of Montrose Road and Hitching Post Lane (AP3323).}
\label{fig-queues}
\end{figure}
\begin{figure}[h!]
\centering
\includegraphics[width=6in]{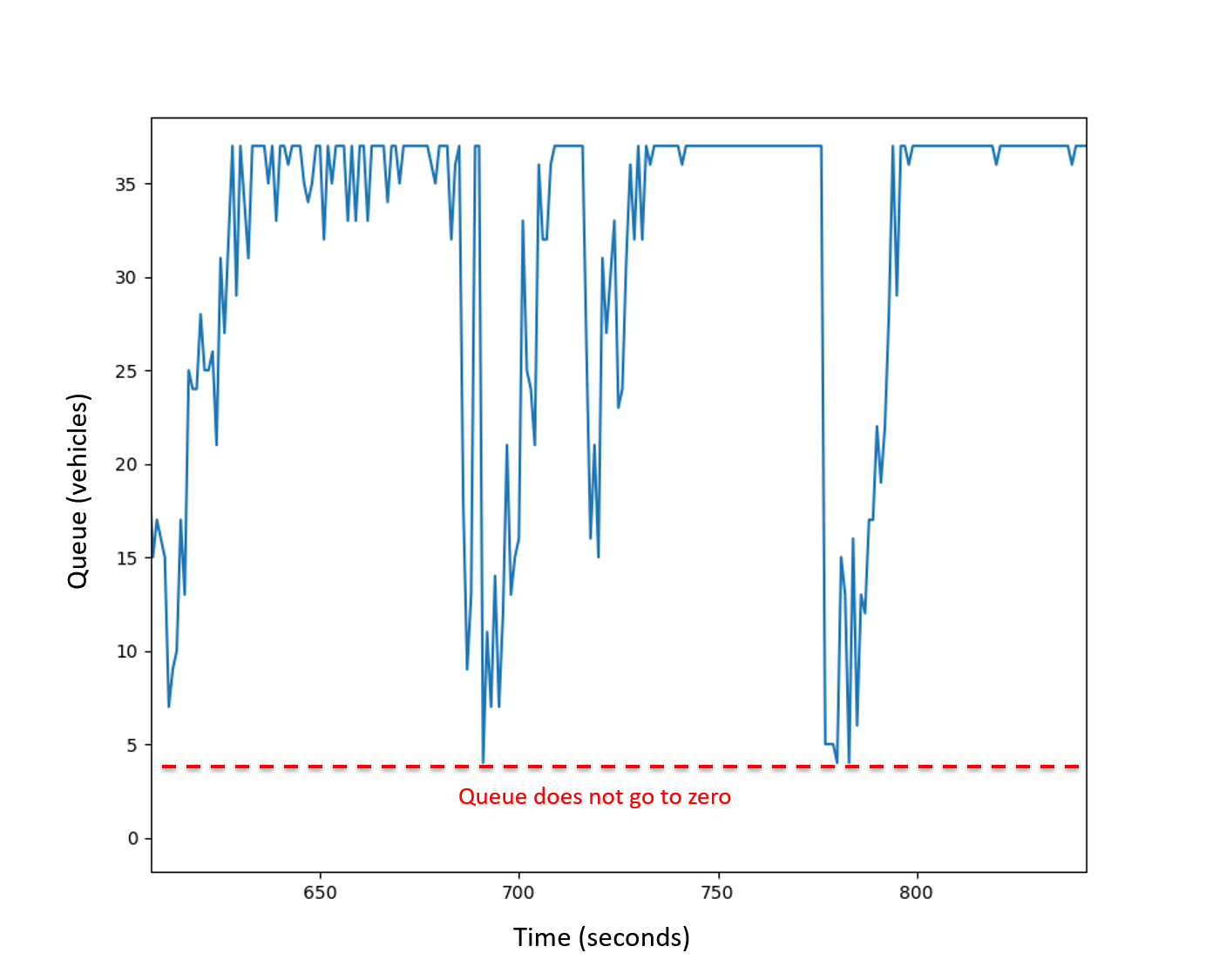}
\caption{Zoomed-in queue dynamics from Figure~\ref{fig-queues} --- queue does not empty.}
\label{fig-queues2}
\end{figure}

General approach to congestion analysis on an arterial network is as follows.
Intersections, where under a given demand a vehicle queue on at least one approach keeps growing
are identified as bottlenecks.
If rearranging the duration of green phases within a cycle leads to a periodic queue behavior --- when
a queue grows then dissolves --- on all intersection approaches,
then this intersection bottleneck is due to poor control.
If, on the other hand, with any phase split we continue observing at least one increasing queue,
then we have a situation of excessive demand.

Figures~\ref{fig-queues}-\ref{fig-queues2} show queue dynamics at the routes A and B approach
to the intersection of Montrose Road and Hitching Post Lane (AP3323).
Vehicle queue measured in SUMO does not grow beyond the storage capacity of a road link, where
this queue is measured.
Once the link storage capacity is reached, queue spills back into the upstream link.
In Figures~\ref{fig-queues}-\ref{fig-queues2}, queue size reaches the storage capacity of 36.
Actually, it grows further, but SUMO reports only the maximum halted vehicles in this
particular link.
What we observe from the queue dynamnics plot, though, is that after a certain time peiod,
this vehicle queue is never fully served.
In other words, number of vehicles in the queue does not go to zero.
It means, that green phase for the vehicles on routes A and B at intersection AP3323
is too short for the given demand.
It so happens that increasing green phase for the movement corresponding to routes A and B
is not a viable option, because that would create a backup on the cross street, Hitching Post Lane.
Thus, we have a case of excessive demand at intersection AP3323.
The problem of excessive demand may be mitigated with decreasing vehicle headways --- through
platoon creation.

We enabled platooning on two links labeled `P' in Figure~\ref{fig-routes} --- upstream and
downstream links of intersection AP3323 on routes A and B.
Then, we ran a series of simulation scenarios varying the fraction of ACC (CACC) vehicles
from 0 to 75\%.
In each simulation two vehicle classes were modeled: ordinary vehicles
and ACC (or CACC) vehicles.
In simulations with CACC vehicles platoons were formed in those two links, where platoning was enabled.
The same number of vehicles was processed in each simulation.
The rates and locations at which cars were generated were identical
in all scenarios to eliminate the variance in randomly generated routes.
For  cases of 0, 25, 50 and 75 percent  ACC (CACC) penetration rate,
we computed average travel time for routes A and B.
Table~\ref{tab-travel-times} lists the resulting mean travel times.

\begin{table}[h]
\centering
\begin{tabular}{|c|c|c|c|c|c|}
\hline
[C]ACC \% & Vehicle Class & \multicolumn{2}{|c|}{Route A} & \multicolumn{2}{|c|}{Route B} \\
\cline{3-6}
 & & ACC & CACC & ACC & CACC \\
\cline{1-6}
0 & ordinary & 711 & 711 & 622 & 622 \\
\cline{1-6}
\multirow{3}{*}{25\%} & ordinary & 696 & 696 & 608 & 607 \\
 & [C]ACC & 691 & 688 & 604 & 598 \\
 & all & 695 & 694 & 607 & 605\\
\cline{1-6}
\multirow{3}{*}{50\%} & ordinary & 671 & 670 & 553 & 547 \\
 & [C]ACC & 665 & 646 & 539 & 523 \\
 & all & 671 & 658 & 546 & 535\\
\cline{1-6}
\multirow{3}{*}{75\%} & ordinary & 661 & 646 & 526 & 516 \\
 & [C]ACC & 660 & 642 & 526 & 509 \\
 & all & 660 & 643 & 526 & 511 \\
\cline{1-6}
\end{tabular}
\caption{Mean travel time in seconds for varying percentage of [C]ACC vehicles on
routes A and B (see Figure~\ref{fig-routes}).}
\label{tab-travel-times}
\end{table}

We can see that ACC vehicles alone (without platooning) reduce the travel time along both routes A and B.
Travel time improvement achieved by CACC over ACC should be attributed to the higher throughput of
intersection AP3323, resulting in smaller queues and thus, smaller waiting times in queues, formed
upstream of this intersection.
Note that ordinary vehicles show reduced travel times, although [C]ACC vehicles have larger gains.
Another observation is that the biggest travel time improvement happens when CACC penetration
rate goes from 25 to 50\%.
The reason is that with 25\% CACC penetration rate, chances that CACC vehicles will be positioned
in sequence, so that a platoon can be formed, are relatively low.
So, CACC case does not show much improvement over ACC with 25\% penetration rate.
On the other hand, when CACC peneration rate is high at 75\%, platoons become much more frequent
and of larger sizes.
This leads to oversaturation of the downstream link, creating a new bottleneck that offsets some of the 
upstream travel time gains achieved by the platooning.

\begin{table}[h]
\centering
\begin{tabular}{|c|c|c|c|c|c|}
\hline
CACC \% & Vehicle Class & \multicolumn{2}{|c|}{Route A} & \multicolumn{2}{|c|}{Route B} \\
\cline{3-6}
 & & Original & Additional Platooning & Original & Additional Platooning \\
\cline{1-6}
\multirow{3}{*}{25\%} & ordinary & 696 & 695 & 607 & 607 \\
 & CACC & 688 & 686  & 598 & 599 \\
 & all & 694 & 693 & 605 & 605\\
\cline{1-6}
\multirow{3}{*}{50\%} & ordinary & 670 & 672 & 547 & 548 \\
 & CACC & 646 & 644 & 523 & 520 \\
 & all & 658 & 658 & 535 & 534 \\
\cline{1-6}
\multirow{3}{*}{75\%} & ordinary & 646 & 646 & 516 & 515 \\
 & CACC & 642 & 641 & 509 & 506 \\
 & all & 643 & 642 & 511 & 508  \\
\cline{1-6}
\end{tabular}
\caption{Comparing mean travel time on routes A and B from the original platooning experiment
with the mean travel time, where platooning is allowed in the additional links, as indicated in
Figure~\ref{fig-routes}.
Travel time is in seconds.}
\label{tab-travel-times2}
\end{table}

To see how platooning can further reduce trave times on routes A and B, we enabled it
on links approaching intersections AP3319 (on route A) and AP3299 (on route B).
Those additional platooning links are labeled `(P)' in Figure~\ref{fig-routes}.
Then, we re-ran simulation scenarios with 25, 50 and 75\% [C]ACC portion of traffic.
Average travel times for routes A and B are summarized in Table~\ref{tab-travel-times2},
comparing them with the results of the original experiment.

As we can see, additional platooning practically does not reduce travel time.
This happens because intersections AP3319 and AP3299 are not bottlenecks,
Pushing vehicles through these intersection in platoons does not
qualitatively change queue dynamics at AP3319 and AP3299 on routes A and B: the green
phase was sufficient to handle unplatooned traffic.

\begin{figure}[h]
\centering
\includegraphics[width=\textwidth]{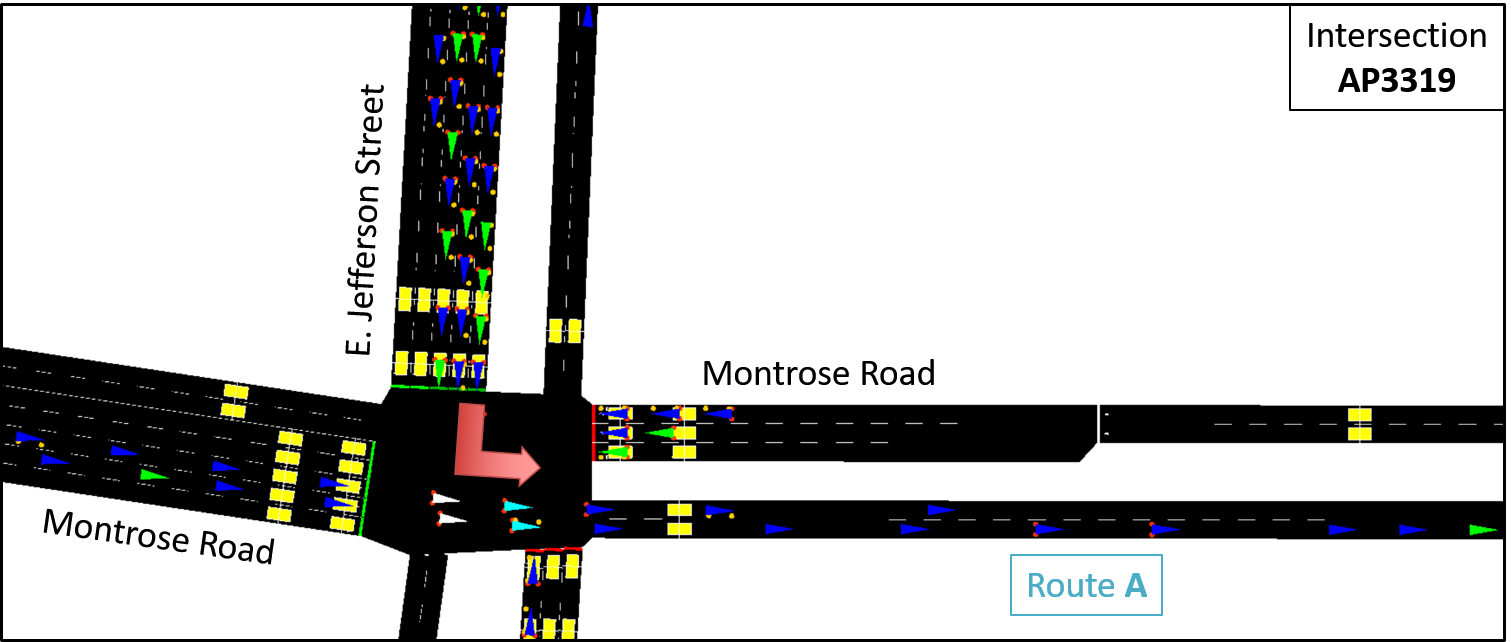}
\caption{Intersection AP3319 --- jam at E. Jefferson Street left turn pocket.}
\label{fig-ap3319}
\end{figure}

Moreover, platooning may cause a problem on cross streets.
In the case of 50\% CACC penetration rate, platooning pushes too much traffic through
intersection AP3319 along route A blocking the link downstream of this intersection for
the left-turning traffic from E. Jefferson Street.
SUMO screenshot depicting this situation is presented in Figure~\ref{fig-ap3319}.

Thus, we can succinctly formulate the platooning rule:
\newline
{\bf platooning should be enabled only at the approaches to bottleneck intersections}.

\section{Conclusion}\label{sec_conclusion}
Analysis of the Gipps, IIDM and Helly car following models shows that:
\begin{itemize}
\item Theoretical bound on intersection throughput defined by the
equilibrium flow may be exceeded in the case of accelerating traffic --- slower
speed is compensated by a larger traffic density.
\item Gipps model exhibits too aggressive acceleration behavior, and,
if used for estimation of intersection throughput, may produce
unrealistically high vehicle counts, which, in the case of free road
downstream, exceed the theoretical bound.
\item Helly model produces deceleration profile that is unacceptable
for drivers and passengers, which makes it not suitable for quantitative
assessment of intersection throughput.
\item Acceleration of IIDM is more gradual than that of Gipps,
and deceleration is more gentle than that of both Gipps and Helly.
IIDM suits better for the analysis of ACC/CACC impact
on the arterial throughput than the other two models.
\end{itemize}

Presence of ACC-enabled vehicles in the traffic increases the
intersection throughput and improves travel time.
The impact of CACC-enabled vehicles that can form platoons largely
depends on how vehicles are ordered in the traffic stream:
CACC vehicles forming a sequence can increase the flow through intersection
significantly more than can be achieved with pure ACC vehicles;
while CACC vehicles interleaved with ordinary ones have the same effect
as pure ACC vehicles.

Ordering of vehicles matters not only in the presence of CACC, but
even just ACC vehicles.
Grouped closer to the head of the queue ACC (CACC) vehicles increase
the intersection throughput more than if they were in the queue's tail.
CACC vehicles interleaved with ordinary ones have the same effect
as  ACC vehicles.

In an urban network the presence of (C)ACC vehicles reduces the queues and hence
the time spent at intersections.
As a result ordinary as well as (C)ACC vehicles benefit from
lower average travel times.

Platooning helps at bottleneck intersections, where bottleneck cannot be dissolved
by changing signal cycle and splits.
Platooning may be harmful in the following cases:
\begin{itemize}
\item downstream intersection is a bottleneck;
\item downstream intersection can become a bottleneck as a result of increased throughput going there;
and
\item flows from other approaches at current intersection can be blocked.
\end{itemize}

\section*{Acknowledgements}\label{sec_acknowledgement}
This research was funded by the National Science Foundation.

\bibliographystyle{plain}
\bibliographystyle{IEEEtran}
\bibliography{IEEEabrv,traffic}

\end{document}